\documentclass[useAMS,usenatbib]{mn2e}
\pdfoutput=1
\usepackage{graphicx}
\usepackage{times}
\usepackage{amssymb}
\usepackage{color}

\newcommand{\abs}[1]{| #1 |}
\newcommand{\nssim}{\mathord{\sim}}

\def\nT{$nT$ }
\def\vap{{\small {VAPOR }}}
\newcommand{\ram}{{\small RAMSES} }

\newcommand{\nut}{{\small NUT} }
\newcommand{\nutco}{{NutCO} }
\newcommand{\msm}{{\small MSM} }

\newcommand{\hmn}{{\small HORIZON-MARENOSTRUM} }

\title[Angular momentum transfer to a Milky Way disk]
{Angular momentum transfer to a Milky Way disk at high redshift}
\author[H.~Tillson et al.]
{H.~Tillson$^{1}$\thanks{email: Henry.Tillson@astro.ox.ac.uk}, 
J.~Devriendt$^{1,2}$, A.~Slyz$^{1}$, L.~Miller$^{1}$ \& C.~Pichon$^{1,3}$\\
$^1$Department of Physics, University of Oxford, The Denys Wilkinson Building, Keble Road, Oxford, 
OX1\,3RH, UK\\
$^2$Centre de Recherche Astrophysique de Lyon, UMR 5574, 9 Avenue Charles Andr$\acute{e}$, 
F69561 Saint Genis Laval, France\\
$^3$Institut d'Astrophysique de Paris, 98 bis boulevard Arago,
F-75014 Paris Cedex, France}

\begin{document}
\pagerange{\pageref{firstpage}--\pageref{lastpage}} \pubyear{2012}
\maketitle
\label{firstpage}

\begin{abstract}
An Adaptive Mesh Refinement cosmological resimulation
is analyzed in order to test whether filamentary flows of cold gas
are responsible for the build-up of angular momentum within a 
Milky Way like disk at $z\geq3$. A set of algorithms is presented
that takes advantage of the high spatial resolution of the simulation 
($12\,\mathrm{pc}$) to identify:
(i) the central gas disk and its plane of orientation;
(ii) the complex individual filament trajectories that connect to the disk, and;
(iii) the infalling satellites.
The results show that two filaments at $z\gtrsim 5.5$, which later merge
to form a single filament at $z\lesssim 4$, drive
the angular momentum and mass budget 
of the disk throughout its evolution, whereas
luminous satellite mergers make negligible fractional contributions.
Combined with the ubiquitous presence of such filaments in {\em all} large-scale cosmological 
simulations that include hydrodynamics, these findings provide strong quantitative evidence 
that the growth of thin disks in haloes with masses below $10^{12}$\,M$_{\sun}$, which host the vast majority of 
galaxies, is supported via inflowing streams of cold gas at intermediate and high redshifts.
\end{abstract}


\section{Introduction}
Our understanding of the origin of angular momentum within galaxies dates
back to the pioneering works of \cite{Hoyle51} and \cite{Peebles69},
who modelled protogalaxies as spherical Eulerian patches
and argued that torques are exerted on a given patch due
to tidal interactions with neighbouring patches. \cite{Doroshkevich70}
removed the assumption of spherical symmetry and 
demonstrated that as perturbations follow
the expansion of the background, their total angular momentum evolves with
the scale factor as $J(a)\propto a^{3/2}$ in an Einstein--de Sitter
universe. An $N$-body 
simulation study by \cite{White84} later confirmed
that this growth of angular momentum continues until the 
perturbation, modelled in a Lagrangian framework, breaks 
away from the background expansion
and collapses to form a halo, whereupon its angular momentum
remains constant. \cite{White84} hence argued that the magnitude of the 
angular momentum acquired for a given halo is set
at the epoch of maximum expansion.

A decade after the seminal papers by \cite{Peebles69} and \cite{Doroshkevich70},  \cite{Fall80}
expanded beyond the early predictions of tidal 
torque theory to account for the hierarchical 
growth of dark matter haloes described in \cite{White_Rees78},
and consequently painted a simple picture of galaxy disk formation.
These authors claimed that gas residing in the potential 
wells of collapsed density perturbations 
is shock heated to the virial temperature of the halo, and 
that the inner gas regions subsequently 
cool and lose their pressure, sinking to the halo centre on a free-fall timescale.
Galaxy disks hence form from the `inside-out',
as the infalling gas retains its specific angular momentum. Several authors
have since reported that the mean stellar age decreases \citep{dejong96,MacArthur04,Gogarten10} 
and specific star formation rates increase \citep{Munoz-Mateos07}
as a function of radial position from the disk centre, 
thereby lending support to the inside-out
paradigm of disk growth.

This standard model of disk galaxy formation makes 
two important assumptions:
(i) the specific angular momentum distributions of the gas and 
virialised dark matter halo are initially equal, and;
(ii) the specific angular momentum of the gas is conserved upon collapse to the central disk.
This simple model has two appealing features. 
Firstly, by making a prediction for the relationship
between the disk scale length and the host halo virial radius,
it is able to recover locally observed distributions of disk 
scale lengths and the $I$-band Tully--Fisher
relation \citep[e.g.][]{Dalcanton97,Mo98,deJong00}.
The second success for this model lies in the discovery, using numerical simulations, of
the remarkable property that the cumulative fraction of mass 
within a halo that has specific angular momentum less than $j$, denoted by $M(<j)$,
is well fit by a universal function that depends on a single free parameter \citep{Bullock01}. This result
holds for both dark matter and (non-radiative) gas present in haloes of all masses, 
although the angular momentum vectors of these two components are not generally aligned \citep{vdb02}.

Despite these successes, recent years have witnessed a resurgence in tackling 
one of the defining characteristics
of the standard theory of disk formation: that gas crossing 
a halo's virial sphere is shock heated to the virial temperature
of the halo, which typically corresponds to X-ray temperatures of a few $10^{6}\,$K. 
\cite{Binney77} was the first to claim that, contrary to the standard paradigm, high density gas
in low mass haloes (i.e. most haloes at high redshift) 
probably crosses an isothermal shock front close to the disk rather than an adiabatic
one at the host halo virial radius, 
and hence argued that the vast majority of gas is expected 
to stream in towards the disk in cold flows with temperatures close to $10^{4}\,$K.
\cite{Birnboim03} reignited this idea by demonstrating that 
for haloes less massive than $\nssim10^{11.6}\,\mathrm{M_{\sun}}$,
gas cools efficiently, loses some of its pressure, and is 
unable to support a virial shock. It was hence concluded
that for haloes of mass $M_{\mathrm{H}} \lesssim 10^{11.6}\,\mathrm{M_{\sun}}$, 
gas flows to the disk in a cold mode as opposed
to a hot mode. The lack of redshift evolution in this threshold mass
and the penetration of cold streams deep inside the virial region of most haloes
was later confirmed by several numerical studies \citep{Keres05,Ocvirk08}.
The existence of a bimodal gas accretion phase potentially has 
enormous implications for star formation, and may, 
as the above authors have speculated, account for the observed lack
of soft X-ray flux and copious Lyman-$\mathrm{\alpha}$ 
emission in high redshift galaxies \citep{Birnboim03},
the bimodality in galaxy colours and its relation to galaxy 
morphology, including the existence of red galaxies at $z>1$
and massive bursts of star formation at $z\sim2\mbox{--}4$ \citep{Dekel06},
and possibly star formation rate downsizing \citep{Keres05,Ocvirk08}. 
These are hence very exciting times for galaxy disk theory.

Having established that gas inflow in most haloes at high redshift proceeds 
via cold flows that often terminate in the outskirts of 
gaseous disks \citep{Brooks09,Powell11}, 
one naturally wonders about the processes governing the transport 
of angular momentum onto disks within these overdense regions.
This is undoubtedly a complex, multiscale problem.
On the largest scales of several Gpc, the Universe is thought
to be arranged in a complex web of cosmic structure 
\citep{Bond96,Pogosyan98}, and simulations
with differing spatial resolution 
have helped to decompose the intricate patterns of this web (e.g. \citealt{Sousbie11}).
It appears that large sheets surrounding Gpc-scale voids
intersect to form filaments that funnel gas from the voids to
the intersection nodes of the web, where haloes
of dark matter form. The amount of mass shared between these separate 
phases has recently been the subject of debate and involves
examining the tidal field tensor that describes the second
order derivatives of the gravitational field, in both the linear
\citep{Doroshkevich70} and non-linear
\citep{Shen06,Hahn07,AragoCalvo10}
regimes of perturbation growth. Yet the general conclusions of these studies
are in agreement: filaments dominate the mass budget.
It is hence likely that this component also carries large 
amounts of angular momentum
to the halo nodes of the cosmic web.

Several recent studies have shed some light on the
subject of angular momentum transport
by filaments on galactic scales.
\cite{Danovich12} performed a statistical study of 
$350$ Milky Way size dark matter haloes selected 
from the \hmn simulation 
(details given in \citealt{Ocvirk08}, \citealt{Dekel09} and \citealt{Devriendt10})
at $z=2.5$ with mass 
$M_{\mathrm{H}}\simeq 10^{12}\,\mathrm{M_{\sun}}$,
whose luminous components were simulated with a physical resolution of $1\,$kpc. 
They found that the streams of cold gas flowing
towards the disk were oriented in a narrow plane, and that the 
angular momentum transported by this infalling component was highly
misaligned with respect to the angular momentum
direction of the disk, until the approximate disk boundary,
whereupon it dramatically swung into close alignment.
An earlier study by \cite{Pichon11} analyzed several outputs from the same run 
and examined the nature of filament trajectories on 
kpc scales, in an attempt to understand the existence
of thin gas disks at high redshift that are thought to 
have formed from the inside-out.
These authors demonstrated that material accreted at the virial
sphere carried more angular momentum at later times,
and attributed this phenomenon to a `lever' mechanism:
recently accreted gas has a larger impact parameter 
owing to the velocity sway of filaments on large scales. 
It was further hypothesized that these large-scale drift velocities
arise from the asymmetric cancellation of motions of gas
pumped out of voids. 
\cite{Pichon11} hence concluded that the
angular momentum transported along cold gas flows
into halo virial regions originates from the asymmetry of surrounding large-scale voids. 
\cite{Kimm11} investigated these claims by probing
an individual halo to a higher physical resolution of $\nssim50\,$pc,
and found that within the virial sphere, 
the specific angular momentum of the radiative gas
was systematically larger than that of the dark matter. 
These authors argued that this was a manifestation of a dark matter 
angular momentum diffusion caused
by the mixing of dark matter particles carrying 
different amounts of angular momentum
at the time of accretion (this mixing is due to the dark matter
undergoing shell-crossing as it passes through walls and filaments, 
\emph{before} entering the host's virial region).
They justified this claim by demonstrating that 
there was a large distribution in the age of the dark matter particles at any 
given radius (see also the impact parameter distributions and backslash velocities 
measured at the virial radius in Figs.\,15 and 20
of \citealt{Aubert07}). On the other hand, the halo gas
does not experience the same angular momentum diffusion
since it streams almost exclusively along filaments (i.e. in a preferential direction) towards the central region
where it eventually experiences an isothermal shock and remains trapped \citep{Birnboim03,Keres05,Ocvirk08}.
\cite{Kimm11} demonstrated that this does indeed happen by showing 
that the amount of specific angular momentum transported by gas
was constant with radius at both low ($0\leq z\leq3$) and high ($z>3$) redshifts, 
except for $r\lesssim0.1r_{\mathrm{vir}}$, whereupon it fell dramatically. 
This hints at unresolved complex dynamics within
the `disk' region (e.g. \citealt{Book11}) of high redshift galaxies. 

Yet due to physical resolution constraints,
the \cite{Kimm11}, \cite{Pichon11} and \cite{Danovich12} studies
were neither able to accurately resolve the 
disk scaleheight and scalelength at high redshift, 
nor the dynamics within the disk region.
By analyzing outputs from one of the high resolution \nut simulations \citep[see][]{Powell11}, which
tracks the high redshift evolution of a single Milky Way like galaxy
with a physical resolution scale of $12\,$pc,
this paper rises to the challenge of 
attempting to account for the amount of angular momentum 
locked-up in a Milky Way like disk that resides
in a halo fed by streams of cold gas, and complements the papers listed above.
The aim is to ascertain whether
filaments dominate the angular momentum budget of the central disk across time.
This would provide quantitative evidence in favour of
the cold mode of disk growth, a theory currently lacking observational confirmation 
\citep[but see \citealt{Fumagalli11}]{Faucher10,Steidel10}.

\vskip1em
This paper is arranged as follows. 
In Section \ref{sim_section}, the details of the simulation 
and the physics encoded within it are briefly discussed.
The various algorithms that have been developed to: 
a) compute the orientation plane of the resimulated Milky Way like disk;
b) resolve the individual gas filaments and satellites that merge onto it, and; 
c) quantify the angular momentum locked-up in each of these galaxy components,
are then introduced in Sections \ref{galaxy_components_methods} and \ref{am_methods}.
A more detailed description of the filament and 
satellite identification algorithms is provided in Appendices \ref{tracer_appendix} and \ref{satellite_appendix}
respectively. Section \ref{results_section} presents the results, Section \ref{discussion_section}
discusses their implications, and Section \ref{conclusions_section} summarizes our conclusions.

\section{The simulation}\label{sim_section}
\subsection{The \nut cooling run}
We examine one of the \nut simulations to understand
how a Milky Way like disk acquires its angular momentum. 
The \nut suite \citep{Powell11} is a set 
of ultra-high resolution simulations that employ
a zoom technique \citep[see e.g.][]{Navarro95} 
to resimulate a Milky Way like galaxy
and follow its evolution across redshift
in a $\Lambda$CDM cosmology. The resolution and number of physical processes that 
are modelled vary from one simulation to the next.
The highest resolution runs in the \nut suite have a maximum spatial resolution
$\Delta x_{\mathrm{res}}$ of $0.5\,$pc but terminate 
at high redshifts ($z\simeq6$) 
due to computational time restrictions, and so in this paper a lower resolution 
$\Delta x_{\mathrm{res}}= 12\,$pc run is 
analyzed as it probes
redshifts down to $z=3$. More specifically, the \nut cooling run, hereafter \nutco run, examined in this paper
has been performed using the publicly available Adaptive Mesh Refinement 
(AMR) code {\small{RAMSES}} \citep{Teyssier02}, incorporating the following physics:

\begin{enumerate}
\item Epoch of reionization---a spatially uniform, redshift-dependent
ultraviolet (UV) background instantaneously switched on at $z=8.5$
(following the \citealt{Haardt96} UV model).
\item Cooling and star formation.
\end{enumerate}

The important technical details of the \nutco run
are now briefly summarized. 
The chosen resimulated Milky Way like halo 
satisfies two conditions: it is a typical 2$\sigma$ density fluctuation, i.e. it is not embedded 
in a denser group or cluster environment at $z=0$,
and its present day mass 
($M_{\mathrm{H}} \simeq 5\times10^{11}\,\mathrm{M_{\sun}}$)
is comparable to the threshold mass 
($M_{\mathrm{H}}\sim4\times10^{11}\,\mathrm{M_{\sun}}$)
reported by \cite{Ocvirk08} below which gas streams along filamentary
structures at low temperatures of T $\lesssim 2\times10^4\,$K.
The entire simulation box has a comoving side length of $9\,h^{-1}\,$Mpc and
starts at $z=499$, with the initial conditions generated
using the package {\small{MPgrafic}} \citep{Prunet08}.
The resimulation box around the main halo has a 
comoving side length of $\nssim2.7\,h^{-1}\,$Mpc, 
and the simulation evolves in a universe of 
WMAP5 cosmology \citep{Dunkley09}
with \mbox{$\Omega_{\mathrm{M}} = 0.258, 
\Omega_{\Lambda} = 0.742$}, \mbox{$\Omega_b = 0.045,
\sigma_8 = 0.8$ and $h = 0.72$}.
The coarse root grid for the entire simulation has $128$ cells along each dimension
of the $9\,h^{-1}\,$Mpc box, whereas the resimulated
region contains three higher resolution nested grids yielding
an equivalent particle resolution of $1024^3$ dark matter particles,
each with mass \mbox{$M_{\mathrm{DM}} = 5.4\times10^{4}\,\mathrm{M_{\sun}}$}.
A quasi-Lagrangian refinement scheme is used to
maintain $\Delta x_{\mathrm{res}} = 12\,$pc in physical coordinates
as the simulation evolves.
Higher levels of refinement are spawned
once either: a) the baryonic mass in the cell exceeds $8m_{\mathrm{SPH}}$
($m_{\mathrm{SPH}} = 9.4\times10^{3}\,\mathrm{M_{\sun}}$) or; b)
the number of dark matter particles in the cell exceeds eight.
This scheme therefore strives to achieve a roughly
equal gas mass per cell.

The \nut simulation includes stars and so a brief discussion of the prescriptions
used for modelling star formation,
which are described in detail
by \cite{Rasera06}, is now provided.
The gas in a cell can cool to temperatures $T\sim10^4\,$K 
via bremsstrahlung radiation (effective until $T\sim10^{6}\,$K), and via
collisional and ionization excitation followed by recombination 
(dominant for $10^4\leq T/\mathrm{K}\leq10^6$). The metallicity of the gas is fixed to $10^{-3}\,\mathrm{Z_{\sun}}$
to allow further cooling to molecular cloud temperatures of 
$\nssim1\,$K via metal line emission, and once the gas is sufficiently dense (i.e. $\rho>\rho_{0}$,
where $\rho_0$ is a density threshold and is chosen to represent the
density of the interstellar medium, hereafter ISM), star formation naturally ensues,
a process which is modelled by a Schmidt law \citep{Schmidt59}. 
The star formation efficiency parameter $\epsilon$ is fixed at $1\%$ 
per free-fall time in concordance with observations
\citep{Krumholz07} and $\rho_0$ is chosen to be 
equal to $400$ hydrogen atoms per cubic centimetre.
If the Jean's scale $\lambda_{\mathrm{J}}$ is close to or below the 
minimum resolved scale $\Delta x_{\mathrm{min}}$ 
(determined by the highest level of refinement $\ell_{\mathrm{max}}$),
it is possible that gas clouds artificially fragment,
leading to artificial star formation. 
Therefore, when in excess of the ISM density $\rho_0$,
the gas is forced to follow
a polytropic equation of state.
The number $N$ of stellar particles that form in a cell whose gas is sufficiently
cool and dense is drawn from a Poisson distribution, based on the cell
mass, the local star formation timescale, and the minimum 
stellar mass allowed for a star particle $m_{\star,\mathrm{min}}$, which itself is proportional
to the minimum cell volume. 
In order to prevent an overproduction of stellar particles in a given cell,
all of the $N$ newly formed stellar particles
are then merged into a single particle of mass $m_{\star} = Nm_{\star,\mathrm{min}}$.
Stellar particles in the \nutco run analyzed in this study are therefore 
stellar clumps of up to several thousand stars.

\subsection{Identifying (sub)haloes \& galaxies and following
their merger histories}

The Most massive Sub-node Method  subhalo-finding algorithm, hereafter {\small{MSM}} \citep{Tweed09},
was used in order to detect the haloes, subhaloes, stellar clumps
and stellar subclumps at each time output of the \nutco run.
\msm assigns a local density estimate to each particle computed using the standard 
Smoothed Particle Hydrodynamics kernel
\citep{Monaghan85}, which weights the mass contributions 
from the $N$ closest neighbouring particles ($N=20,32$ for the haloes
and stellar clumps, respectively).
Haloes are then resolved by imposing a density threshold criterion and by measuring local
density gradients. In order to detect host haloes, stellar clumps and associated 
levels of substructure, \msm successively raises the density
thresholds on the host until all of its node structure has been resolved.
The most massive leaf is then collapsed along the 
node tree structure to resolve the main halo or the main stellar clump
and the same process is repeated for the lower mass leaves, defining the embedded substructures. 

Each resolved halo (stellar clump) object is forced to contain at least
$40$ ($100$) particles to ensure reliable detections.
Note that the stellar clump resolution limit is set above the dark halo limit to avoid
identifying groups of stars within the disk as independent objects.
The TreeMaker code \citep{Tweed09} 
is then used to link together all the time outputs by
finding the fathers and sons of every halo and subhalo,
and every stellar clump and subclump.

\section{Resolving the galaxy components}\label{galaxy_components_methods}

This section describes the algorithms that have been used to identify the
likely sources of the disk's angular momentum within the host's virial region.
The following terms---$r_{\mathrm{vir}}$,
disk, satellites and virial sphere---are henceforth made in reference to the \nut host halo.

\begin{figure*}
\centering
\includegraphics[width=0.85\columnwidth,height=7cm]{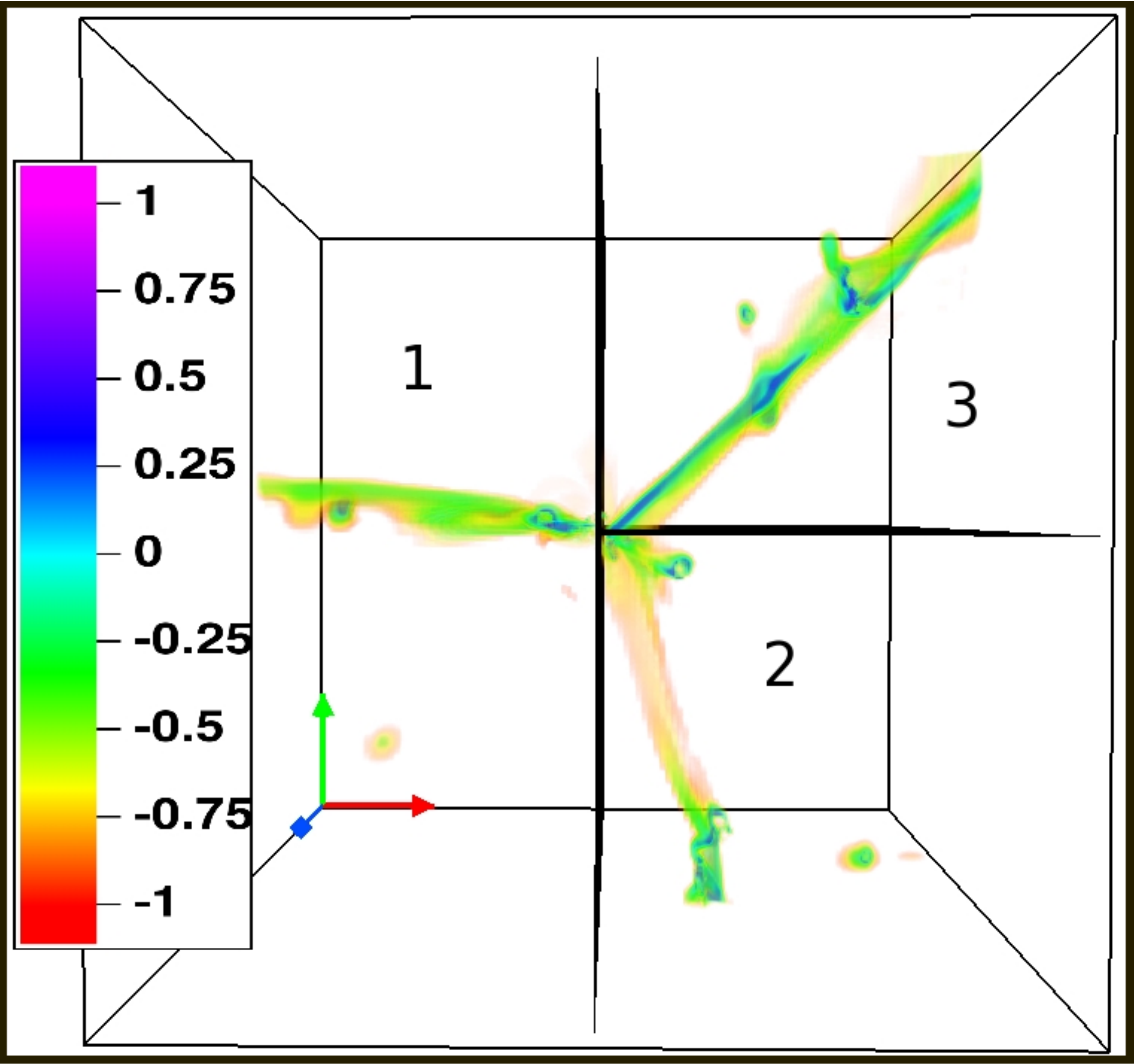}
\includegraphics[width=0.85\columnwidth,height=7cm]{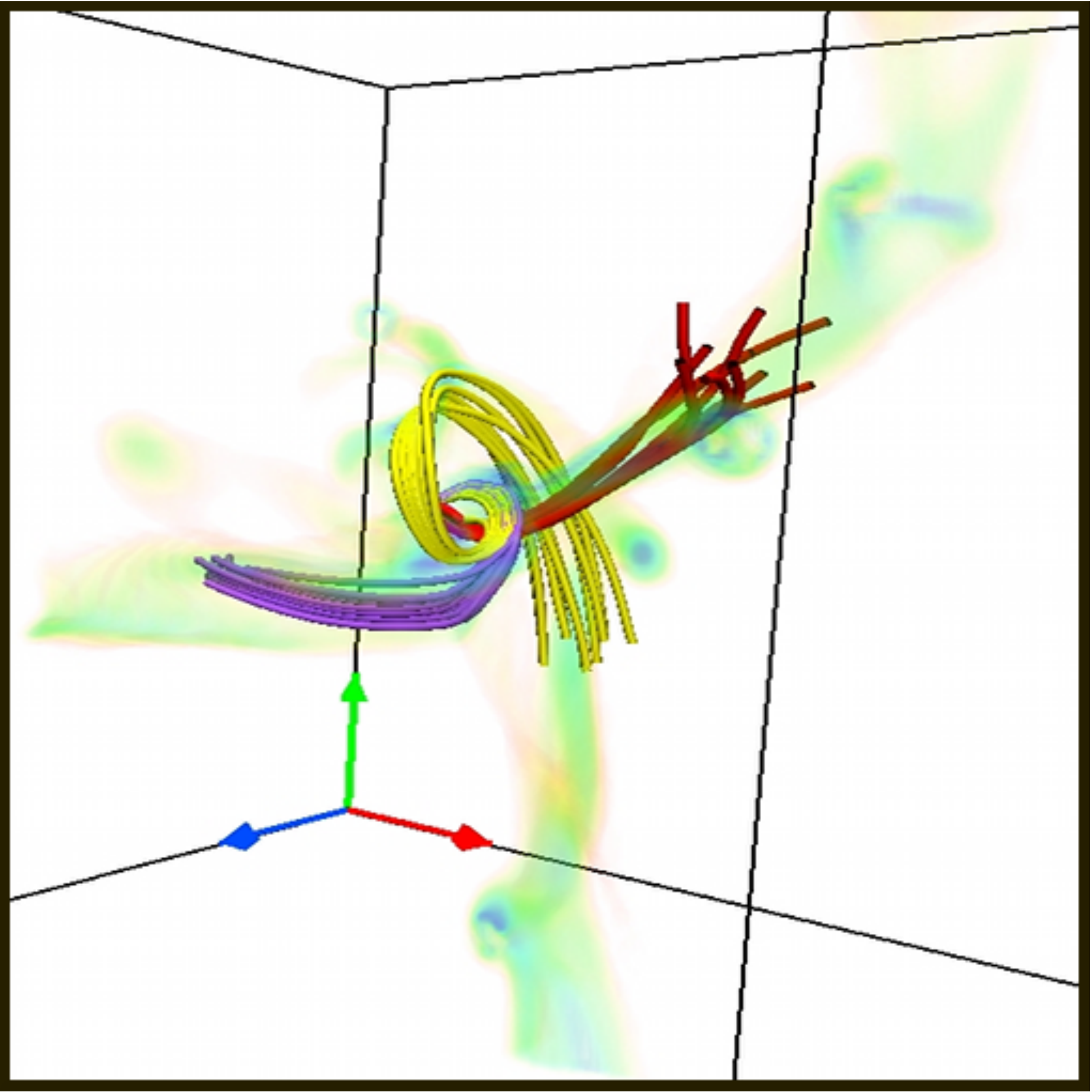}
\caption{Images of the filament gas within a box centred on the main host
at $z\sim10$ (left) and $z\sim8$ (right) in the \nutco run, generated 
using the Visualization and Analysis Platform for Ocean, Atmosphere, and Solar Researchers 
software (commonly referred to as {\small{VAPOR}}, see {\tt https://www.vapor.ucar.edu/} for details). The sidelength of the box
in the left panel corresponds to $4r_{\mathrm{vir}}$ ($\nssim 17\,$kpc) 
and this face-on view illustrates the
clear separation of the filaments into three distinct regions 
(labelled accordingly) at large scales and early times.
The image in the right panel shows the inner virial region ($\nssim 7.5\,$kpc) 
at lower redshift when the dynamics
of the system have changed,
and the velocity flow tubes demonstrate the 
complex trajectories of gas flow within the
gas disk region, colour coded according to the filament that they are likely to trace. The density
of the gas has been dimmed and the viewing angle rotated to best highlight each trajectory. 
Both panels have been refined to $\ell=16$, 
corresponding to a physical cell width of $\nssim 17\,$pc (left) and $\nssim 22\,$pc (right),
and the colour bar indicates the gas density
in dimensionless units of $\mathrm{log_{10}(n/cm^{-3})}$,
where $n$ is the number density of hydrogen atoms per cubic centimetre.
Filament gas has been detected by applying the \nT criteria,
but gas with density $\mathrm{log_{10}(n/cm^{-3})} < -1$
has been excluded from both figures for clarity.
The rightward red, upward green and forward blue axes
correspond to the $x$, $y$ and $z$ directions, and
the vertical and horizontal black planes in the left panel show the spatial
cuts in the $x$ and $y$ directions that have been used to start the 
colouring of the tracer particles, which is 
a technique that has been developed to separate each of the filaments 
at high redshift (see Section \ref{tracer_section} and Appendix \ref{tracer_appendix}).}
\label{filaments_visualization}
\end{figure*}

\subsection{Grid notations}\label{section_grid_notations}

This study analyzes cubic and spherical grids $\mathcal{G}_x$ 
centred on the galaxy (i.e. on its most dense cell) with physical half-lengths and radii equal to $xr_{\mathrm{vir}}$, 
where $x\in\{0.1,0.15,0.5,1.0,2.0\}$.
The numerical subscript $x$ will be explicitly stated when discussing a certain grid type, 
but for comparisons between grid types 
(e.g. the gas grid $\mathcal{G}_{g}$ or the stellar grid $\mathcal{G}_{\star}$)
it is replaced by a symbol denoting the type, 
and the extent of the grid is implicitly understood.
All of the spherical (cubic) grids in this paper 
with physical diameter (full-length) $L$ are subdivided into $n$ cubic cells of 
equal physical sidelength $\Delta x$ and are hence `fixed', obeying the relation:
\begin{equation}
\Delta x = \frac{L}{n}.
\label{eqn_deltax}
\end{equation}
Since there are $2^{\ell}$ cells
along each dimension of the physical simulation grid length $L_s(z)$ 
at level $\ell$, it follows that
the refinement level of $\mathcal{G}_x$ is given by:
\begin{equation}
\ell = \frac{\ln\left(L_s/\Delta x\right)}{\ln\,2},
\label{eqn_ell}
\end{equation}
where $L_s(z) = 9\,h^{-1}\,\mathrm{Mpc}/(1+z)$ and $\ell$ is rounded to
its nearest integer value.
The routines
that generate fixed 
density, temperature and velocity
grids at a given location by performing an interpolation
from the simulation grid for gas properties and a Cloud-in-Cell
smoothing technique for particle properties,
require $L$ and $\ell$ as inputs at runtime. 
Memory and CPU time constraints mean that $n$ is not allowed to exceed $1024$, however,
so we decrease $\ell$ successively in integer units in equation (\ref{eqn_ell}) until this constraint is satisfied. 
As a result, the resolution $\Delta x$ for the grids enclosing 
the $r_{\mathrm{vir}}$ and $2r_{\mathrm{vir}}$ regions
deviates from the maximum resolution $\Delta x_{\mathrm{res}}$ for low redshift outputs,
as the virial radius of the host (and hence $L$) expands. 
However, this `time bias' does not affect the grids $\mathcal{G}_{0.1}$ and $\mathcal{G}_{0.15}$
used for performing the angular momentum computations (see Section \ref{am_methods}), 
as their cell width remains equal to $\Delta x_{\mathrm{res}}$ at all times.

\subsection{Disentangling the filaments using a tracer particle colouring algorithm}\label{tracer_section}

\cite{Powell11} have shown that parcels of gas whose number density 
of hydrogen atoms $n$ and temperature $T$ 
simultaneously satisfy $0.1\leq n/\mathrm{cm}^{-3}\leq10$ and 
$T/\mathrm{K}\leq2\times10^4$, are representative of gas
belonging to filaments in the \nut simulations at high redshift ($z\geq9$).
The analysis in this work resolves the \nut filaments down to $z=3$, 
hence in order to tackle the possible time evolution in their density, 
this paper uses the
prescription between the lower threshold
and the background density introduced by \cite{Kimm11}, who argued that
large-scale filament density is gravitationally coupled to the expansion
of the Universe. They defined the lower number density bound $n_{\mathrm{L}}$ as:
\begin{equation}
n_{\mathrm{L}} \equiv \frac{\delta_f \bar{\rho} f_b X_{\mathrm{H}}}{m_{\mathrm{H}}},
\label{eqn_kimm}
\end{equation}
where $\bar{\rho}, \delta_f, f_b, X_{\mathrm{H}}$ and $m_{\mathrm{H}}$
are respectively the mean density of the background,
the density contrast of the filament (measured with respect to $\bar{\rho}$),
the fraction of the total mass in baryons ($\equiv \Omega_{b}/\Omega_{\mathrm{M}}$),
the primordial relative mass abundance of hydrogen ($76\%$), and the mass of a hydrogen atom. 
A typical value of $n_{\mathrm{L}}$ from the \nutco run at $z=3$ is $\nssim0.01\,\mathrm{cm^{-3}}$.

\begin{figure*}
\centering
\includegraphics[width=\columnwidth,height=6.5cm]{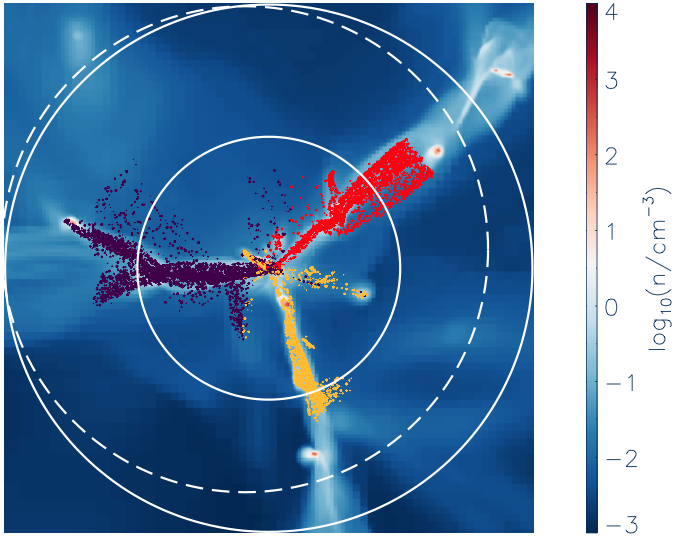}
\includegraphics[width=\columnwidth,height=6.5cm]{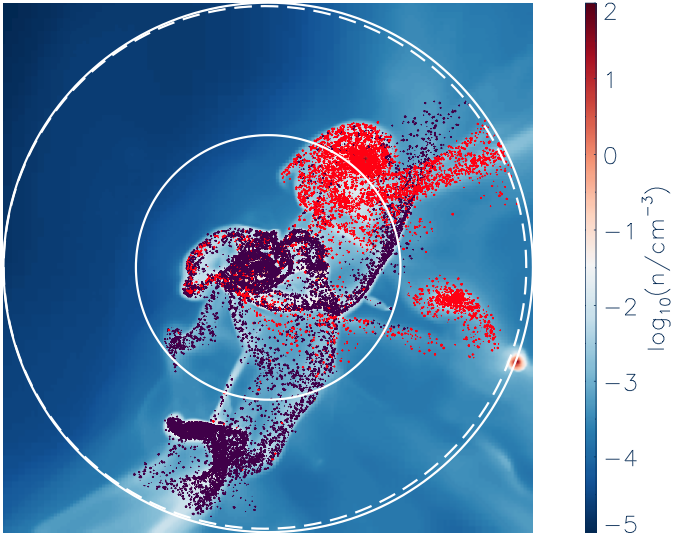}
\caption{These gas density images are $z$ projections of the large-scale region
surrounding the central disk in the \nutco run, and illustrate
the spatial distribution of particles tracing the filaments
at $z\sim9$ (left panel) and $z=3$ (right panel). 
The inner and outer solid circles have
radii equal to $r_{\mathrm{vir}}$ and $2r_{\mathrm{vir}}$ respectively,
where $r_{\mathrm{vir}}\sim5.3\,$kpc (left) and $r_{\mathrm{vir}}\sim31\,$kpc (right).
The dashed circles represent the $2r_{\mathrm{vir}}$
regions at the previous time output, corresponding to $z\sim9.5$ 
($r_{\mathrm{vir}}\sim4.7\,$kpc) and $z\sim3.04$ ($r_{\mathrm{vir}}\sim30\,$kpc).
All of the circles are centred at the location of the densest cell in the disk region
at the relevant time output.
Left panel: Those tracer particles that have been identified with each of the purple, yellow and red
filaments at $z\sim9.5$ and that coincide with filament cells refined to $\ell=15$ ($\Delta x\sim38\,$pc)
at $z\sim9$, have been updated
to their new positions.
The offset in centres between
the outer solid and dashed circles
arises due to the motion of the centre of disk rotation,
and the gap between each set of coloured tracers and the dashed circle
highlights the displacement of the tracers over the time interval.
Uncoloured new tracer particles that have crossed the $2r_{\mathrm{vir}}$ sphere over the interval
and that move along filament trajectories
at $z\sim9$ have been omitted.
Right panel: The filament configuration at $\ell=14$ ($\Delta x\sim190\,$pc) 
for the final time output at $z=3$. 
This image includes both updated coloured tracers 
and new tracers that have been assigned a colour using the nearest neighbours method
(see text). In Section \ref{results_section} it is demonstrated that the purple and yellow filaments start
to merge around $z\sim5.5$, with the former filament surviving,
and this is confirmed by the absence of yellow tracers in this panel. 
Only $5\%$ of all the coloured filament tracers are shown at this time output, for clarity.}
\label{distr_tracer}
\end{figure*}

It was also found that an upper temperature limit of $T_u\leq2\times10^5\,$K (e.g. \citealt{Keres09,Faucher11})
detected filaments to a higher level of accuracy
for $z\geq3$ in the \nutco run than the classical
cold mode temperature of $T_u\leq2\times10^4\,$K (e.g. \citealt{Kay00,Birnboim03,Keres05})
at which gas is thought to cool via Lyman-$\alpha$
emission (e.g. \citealt{Fardal01}). There
are two reasons for this. Firstly, once reionization has completed 
($z_{\mathrm{re}}=8.5$), the UV background heats all the gas 
uniformly, without taking into account its capacity to self-shield from UV photons in denser environments.
Secondly, the absence of metal production sources in the \nutco run
causes the gas metallicity to remain fixed at its very low initial value (10$^{-3}$ Z$_{\sun}$). 
As a consequence, cooling is inefficient and filaments stay hotter for longer (e.g. \citealt{Sutherland93}).

\begin{figure*}
\centering
\includegraphics[width=1\columnwidth,height=7cm]{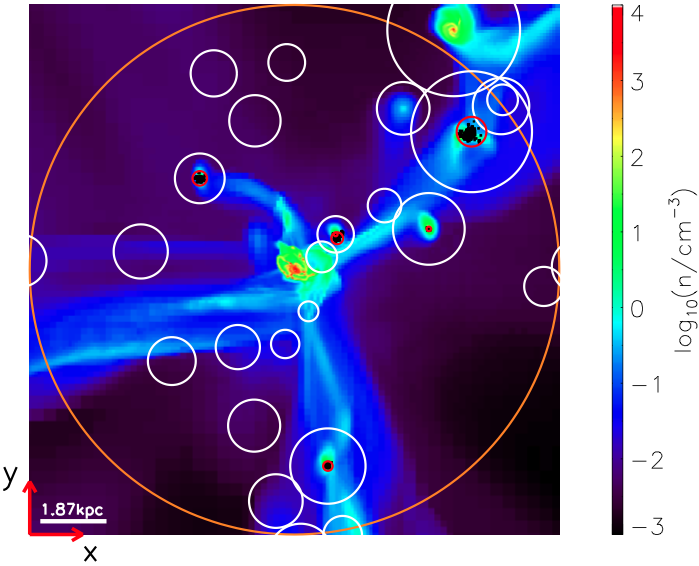}
\includegraphics[width=1\columnwidth,height=7cm]{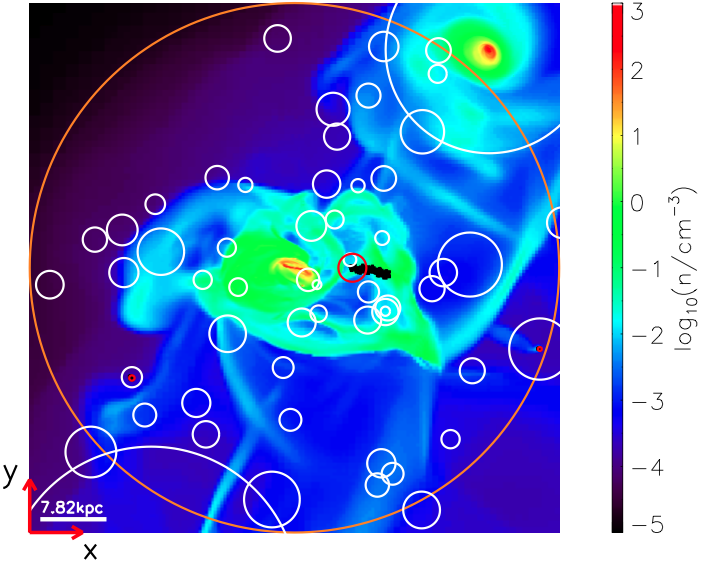}
\caption{Projections along the $z$ direction of all the gas within a box of 
sidelength equal to $2r_{\mathrm{vir}}$,
corresponding to the diameter of the orange circles.
The centre of each image coincides with the centre of
the stellar clump of the host galaxy at 
$z\sim8$ (left) and $z=3$ (right), with
$\ell=16$ ($\Delta x\sim22\,$pc)
and $\ell=15$ ($\Delta x\sim95\,$pc).
The virial regions of the stellar and dark halo satellites of the main host, as identified by 
the satellite-finding algorithm, are shown by the red and white circles respectively,
with each circle centred at the densest member star identified as part of the satellite. 
The constituent satellite stars have been detected by the clump-finder,
and are shown as small filled black circles.
The white horizontal bars
and vertical colour bars indicate the length scale
(in physical kpc) and the
density colour scale (in dimensionless units of $\mathrm{log_{10}}$ number of hydrogen atoms
per cubic centimetre) respectively. 
Note that the large stellar satellites
in the upper right corners of both panels have not been identified
as luminous satellites because their stellar virial regions do not infringe the host's
virial sphere at these epochs. }
\label{satellites_im}
\end{figure*}

Hence, in order to identify the neutral hydrogen and helium 
gas in the filamentary phase in the \nutco run, 
the density and temperature criteria, hereafter $nT$,
reported by \cite{Powell11} have been used for $z\geq 3$,
but the lower density threshold has been replaced by the value
given by equation (\ref{eqn_kimm}) and the upper 
temperature limit has been increased to $T_u=2\times10^5\,$K.
The left panel in Fig.~\ref{filaments_visualization} shows an image 
of the three filaments within $2r_{\mathrm{vir}}$ of the host centre at $z\sim10$ in the \nutco run,
found by imposing these \nT criteria.
The filaments occupy three distinct regions 
on large scales at this epoch and the flow appears to be ordered
and mostly radial (a discussion of 
how gas that is shared by filaments and satellites is apportioned 
between these two phases is preserved for Section \ref{satellite_section}). 
The dynamics of the gaseous motions in
the central region are far more complex, however, 
as demonstrated in the right panel,
which shows the velocity flow vectors of filament gas
within the virial region at a later time (corresponding to $z\sim8$) when the system 
has evolved to a different configuration. 
The individual filament trajectories bend, twist
and mix with each other in the central region, yet often remain intact. In order 
to test the hypothesis that the disk acquires its angular momentum
from filamentary motions, it is critical to follow these separate individual filament 
trajectories. A suitable tool for performing
this Lagrangian exercise is the use of tracer particles \citep[e.g.][]{Dubois12}, which have been demonstrated to preferentially 
trace filament flows in galaxy simulations \citep{Pichon11}. 
The tracer particles in the \nut  suite have an identical spatial distribution to dark matter
at the beginning of the simulation, and are assigned zero mass. 
They are simply advected with the gas and can therefore be used to track its velocity 
field within the host virial region.

The algorithm that resolves each filament 
at high redshift is henceforth referred to as the `tracer propagation
method', and is described in detail in Appendix \ref{tracer_appendix},
which also includes a discussion of the (negligible) uncertainties associated with the method.
The main idea is to assign the tracer particles
within each filament a unique colour at some early time, 
and then use a nearest neighbours scheme to construct the individual
filament trajectories within the host virial region at later times.
Throughout this paper reference is made
to the purple, yellow and red filaments in regions $1, 2$,
and $3$ of the left panel of Fig.~\ref{filaments_visualization} respectively.
The left panel of Fig.~\ref{distr_tracer} shows tracer 
particles colour coded according to the filament
they sample at $z\sim9.5$, updated to their new filament positions at $z\sim9$.
The displaced tracers extend across the full virial region 
at $z\sim9$ (given by the inner solid circle), 
and so new tracers entering the $2r_{\mathrm{vir}}$ region at this epoch
(denoted by the outer solid circle) are appropriately coloured.
Since colour is propagated across time,
one is able to test the overall performance of the method by considering the final output. 
Coloured filament tracers within $2r_{\mathrm{vir}}$ at $z=3$ are hence
shown in the right panel of Fig.~\ref{distr_tracer}
(only $5\%$ of all the coloured filament tracers are plotted for clarity, 
but this subsample is representative of the whole distribution). The purple
and yellow filaments undergo a merger around $z\simeq5.5$ in the simulation 
(this is discussed in more detail in Section \ref{results_section}), and the 
algorithm clearly captures the aftermath of this merger as
only the purple and red tracer particles remain at $z=3$. The filaments are highly 
perturbed within the inner virial region and trace
complex trajectories, no longer displaying the ordered
gas inflow seen at higher redshift in the left panel. 

\subsection{Locating the satellites}\label{satellite_section}

There are two types of satellites in the simulation:
ones with a dark matter halo and ones whose dark halo component
has been stripped by tidal forces to the extent that it does not contain enough particles
to be detected by the halo finder.
One property all satellites share in common, however, is that they are substructures 
of the main host. As shown in the left panel of Fig.~\ref{satellites_im},
luminous satellites in the simulation are
often found to stream along filaments, 
eventually merging onto the central disk,
perhaps transporting large amounts of angular momentum in the process.
One is therefore faced with the following dilemma:
should the angular momentum of gas gravitationally bound
to a satellite, which itself is drifting
along a filament, be solely attributed to the satellite?
The answer to this question in this study is chosen to be `no':
only gaseous material within the virial region of a satellite
that does not satisfy the \nT criteria in that
it lies above the upper filament density cut,
is apportioned to the satellite phase.
`Satellite' is therefore made in reference to the central dense galaxy component
which dominates the `true' satellite angular momentum signal. 
The advantages of this satellite definition are two-fold:
\begin{itemize}
\item only small regions are masked from the density field,
leaving the filament trajectories largely unperturbed, and;
\item the highest density material that is likely to dominate
the satellite angular momentum budget is accounted for.
\end{itemize}

The satellite identification procedure used
in this study locates subhaloes of the main host
that were once themselves field haloes earlier in their accretion history.
Fig.~\ref{satellites_im} shows $z$ projections of the 
gas density within the host virial region (orange circle) at $z\sim8$ (left panel) and $z=3$ (right panel), 
and includes all the dark matter satellites
(white circles) and stellar satellites (red circles) that have been
found by applying the satellite identification technique,
which is described in detail in Appendix \ref{satellite_appendix}.
Each of these circles is centred on the centre of the galaxy it represents, 
as identified by the halo and stellar clump-finder.
The filled black circles correspond to the
constituent stellar particles of each stellar satellite,
which can violate the virial accuracy condition imposed
by the \msm halo-finder and can hence lie beyond their host's virial region,
as illustrated for the elongated satellite in the right panel.
Clearly the apparent luminous satellites in Fig.~\ref{satellites_im}
are captured by the satellite-finding algorithm 
at the two extrema of the range in epochs examined in this study. 

\section{Angular momentum measurement techniques}\label{am_methods}

\subsection{Calculating angular momentum in the general case}

Unless stated otherwise, the total angular momentum $\bmath{J}$ has been computed
by using the following expression:
\begin{equation}
\bmath{J} = \sum_i m_i{(\bmath{r}_i - \bmath{r}_c) \times (\bmath{v}_i - \bmath{v}_c)}\label{J_eqn},
\end{equation}
where $i$ refers to either a stellar particle, a dark matter particle
or a gas cell of mass $m_i$ at location $\bmath{r}_i$ with velocity $\bmath{v}_i$,
and $\bmath{r}_c$ and velocity $\bmath{v}_c$ 
refer to the centre of rotation within the central galaxy. There are two
suitable estimators for these latter quantities: 
1) the densest cell, which is likely to correspond to the disk centre, and; 
2) the centre of mass.
Both of these definitions have been found to yield
very similar estimates of $\bmath{r}_c$ and $\bmath{v}_c$ in the \nutco run, but
at high redshifts where the main host undergoes major mergers, 
it sometimes happens that the centre of mass is
located in a region relatively devoid of mass, which does not correspond to the 
centre of the disk. We therefore adopt
the densest cell as the best estimate of $\bmath{r}_c$, which
has been computed by simply summing 
the gas $\mathcal{G}_g$, stellar $\mathcal{G}_{\star}$ and dark matter $\mathcal{G}_{\mathrm{DM}}$ 
grids within $0.1r_{\mathrm{vir}}$ of the host centre and locating the member cell 
of their sum $\mathcal{G}_{0.1}$ with the most mass.
The velocity of the centre of disk rotation $\bmath{v}_c$ has been approximated by
the centre of mass velocity $\bmath{v}_{\mathrm{com}}$ of $\mathcal{G}_{0.1}$. Since
there are a large ($\gg 10^6$) number of cells and particles that contribute to $\bmath{v}_{\mathrm{com}}$
at all times, noise is low and this velocity should thus be very close to the true bulk velocity of the disk.

\subsection{Obtaining the equation of the gas disk's plane of rotation}\label{disk_plane_section}

\begin{figure}
\centering
\includegraphics[width=\columnwidth,height=6cm]{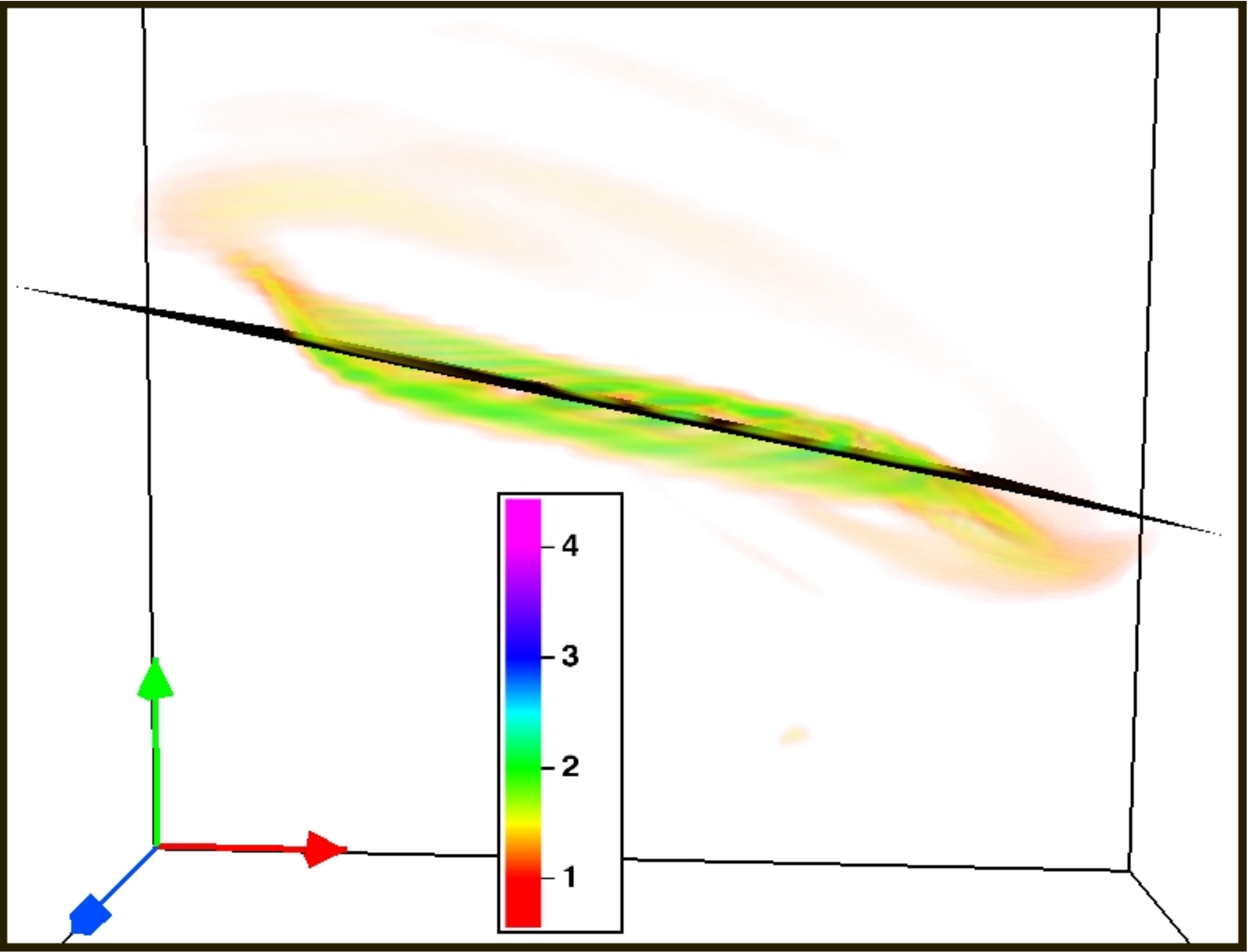}
\caption{A \vap visualization in density of the gas disk refined to $\ell=18$ ($\Delta x\sim12\,$pc)
at $z=3$. The black slice shows the plane of disk rotation found by computing the disk
angular momentum $\bmath{J}_d$ within $\mathcal{G}_{0.1}$
\mbox{(equation \ref{J_eqn})}. The scaled sidelength of the box in this image is approximately
$6\,$kpc in physical coordinates, and the set of orthogonal axes and the units of the colour bar are
the same as in Fig.~\ref{filaments_visualization}.}
\label{disk_visualization}
\end{figure}

It is assumed that the rotation of the gas disk is
confined to a single plane centred on $\bmath{r}_c$ at all times,
and an estimate for the equation of this plane has been 
made by using equation (\ref{J_eqn}) to compute
the disk angular momentum $\bmath{J}_d$. All of the gas within
$\mathcal{G}_{0.1}$ associated with the luminous satellites
(identified using the satellite-finding technique) and the filaments (identified using the \nT criteria)
has been flagged, and the remaining gas in this region is assigned to the disk. 
Fig.~\ref{disk_visualization} shows a density image of the 
gas disk at $z=3$ in the \nutco run, and overplots the plane
found by adopting the above disk definition.
The excellent agreement between the computed disk plane 
and its apparent orientation in Fig.~\ref{disk_visualization} provides 
confirmation that the disk definition and estimates of $\bmath{r}_c$ and $\bmath{v}_c$ 
in equation (\ref{J_eqn}) are reliable.

\subsection{Quantifying angular momentum transport to the disk}\label{flux_section}

The material channelled inward along filaments $f$ in the simulation gets compressed
once it settles into the disk $d$, and star formation ensues when the gas is sufficiently dense.
Mass and angular momentum transported by these flows is hence assumed to be shared
amongst the central baryons $b$, as shown below by the arrows.
The central baryons refer to both the gas in the disk and the stars that 
form from the disk's fragmentation:
\begin{eqnarray}
m_f \to m_b &=& m_d + m_{\star}\label{eqn_disk_mass}\\
\bmath{J}_f \to \bmath{J}_b &=& \bmath{J}_{d} + \bmath{J}_{\star}.\label{eqn_disk_am}
\end{eqnarray}
At a given time output, a list
of identifiers of every particle that has ever 
been resolved as a member constituent of a satellite, 
as detected by the satellite-finding algorithm over the Milky Way host's virial region, is updated. 
Hence $\bmath{J}_{\star}$ has been computed at each epoch by using equation (\ref{J_eqn}) 
and including each particle within $0.1r_{\mathrm{vir}}$
that does not belong to this list. 
These non-satellite stars correspond to bulge and disk stars, but the former
component has a weak angular momentum modulus (see e.g. \citealt{vdb02})
and so the stellar disk signal dominates $\bmath{J}_{\star}$. 
Hence, even though $m_b$ and $\bmath{J}_b$ refer to the total (stars plus gas) mass and angular momentum of
the `baryonic galaxy' in what follows, we will often abbreviate this to `the disk'. By
excluding contributions from satellite stars, 
equations (\ref{eqn_disk_mass}) and (\ref{eqn_disk_am}) focus
on \emph{in situ} disk formation.

We now compare $\bmath{J}_b$ at each time output with the accumulated angular momentum
brought in by the satellite and cold filament gas 
within $0.1r_{\mathrm{vir}}$, since these components are thought to fuel the disk's
angular momentum budget. We also estimate
the fraction of $\bmath{J}_b$ acquired via a hot mode of gas accretion, 
as this enables direct comparisons with the cold mode signals.
This hot gas phase is populated by selecting all of the gas cells with a
temperature above the imposed upper temperature limit of the cold phase (i.e. $T>2\times10^5\,$K).
Hot gas should not be regarded as a separate component, however, because these cells
are included in the satellite and disk signals by construction
(see Section \ref{disk_plane_section}). 

\subsubsection{The hot and cold phase contributions to $m_b$ and $\bmath{J}_b$}\label{shell_section}

Estimates of the angular momentum accumulated onto the disk via 
the accretion of hot and cold gas
have been made by time-integrating 
the instantaneous flow rates of these components
across a thin spherical shell located on the 
periphery of the disk region. This is a suitable method to employ 
because the inflowing gas in the simulation settles onto the disk
very rapidly after crossing the shell (the validity of this assumption is tested
in the following section). The flow rates have been computed by using
the technique described in \cite{Powell11},
and for simplicity, reference is made to the filament gas phase in the descriptions below
(but the same equations apply to the hot mode):
\begin{eqnarray}
\frac{\mathrm{d}\bmath{J}}{\mathrm{d}t} &=& \frac{\sum_i \rho_i v_{r,i}\left[(\bmath{r}_i-\bmath{r}_c)
\times(\bmath{v}_i-\bmath{v}_c)\right](\Delta x)^3}{\sum_j(\Delta x)^3}4\pi r_m^2\label{eqn_am_flux}\\
\frac{\mathrm{d}M}{\mathrm{d}t} &=&  \frac{\sum_i \rho_i v_{r,i}(\Delta x)^3}
{\sum_j(\Delta x)^3}4\pi r_m^2.\label{eqn_mass_flux}
\end{eqnarray}
The sum indexed by $j$ is performed
over all of the cells in the shell, irrespective of their gas phase,
and $\bmath{r}_c$ and $\bmath{v}_c$ are the same as in equation (\ref{J_eqn}).
The variables $\bmath{r}_i, \bmath{v}_i$ and $\rho_i$ 
correspond respectively to the position, velocity and
density of the $i^{\mathrm{th}}$ cell satisfying the relevant imposed condition,
which in this case is the requirement that $i$ is a filament cell.
Equations (\ref{eqn_am_flux}) and (\ref{eqn_mass_flux})
hence yield the net radial flux transported 
across the spherical shell towards (negative sign) or away from (positive sign) the disk,
where the radial velocity of each cell is given by:
\begin{equation}
v_{r,i} = (\bmath{v}_i-\bmath{v}_c)\cdot\frac{\bmath{r}_i-\bmath{r}_c}{|\bmath{r}_i-\bmath{r}_c|}.
\label{eqn_vr}
\end{equation}
It follows that the total angular momentum (and mass)
transferred towards the disk by each filament over the time interval $\Delta t = t - t_i$ is:
\begin{eqnarray}
\Delta \bmath{J}_{\mathrm{flux}}(t)&=& \int_{t_i}^{t} \frac{\mathrm{d}\bmath{J}}{\mathrm{d}t}\,\mathrm{d}t\\
&\simeq& \left(\frac{\mathrm{d}\bmath{J}}{\mathrm{d}t}\bigg |_t + 
\frac{\mathrm{d}\bmath{J}}{\mathrm{d}t}\bigg |_{t_i}\right)\frac{\Delta t}{2},\label{eqn_trap_int}
\end{eqnarray}
where the latter step approximates the integral by 
a simple numerical trapezium integration,
i.e. a linear interpolation of the unresolved
flow rates between time outputs.
The signal found using the `spherical flux technique' of equation (\ref{eqn_trap_int}) 
at $t$ is then accumulated with
the corresponding signals at earlier times.

The summations indexed by $i$ in equations (\ref{eqn_am_flux}) and (\ref{eqn_mass_flux}) 
include all of the filament cells whose centres
lie within a shell of inner, outer and mid radius
$r_{\mathrm{in}}=r_m-\Delta r/2$, $r_{\mathrm{out}}=r_m+\Delta r/2$ and $r_m=0.1r_{\mathrm{vir}}$, respectively.
The thickness of the shell $\Delta r$ is a free parameter and has 
been fixed to $1\%$ of $r_{\mathrm{vir}}$ (i.e. one tenth of the $0.1r_{\mathrm{vir}}$ radius)
at each time output, as this choice strikes a good compromise between
negligible Poisson error and the validity of the thin shell approximation
(the shells typically included $\nssim2\times10^5$ cells at the highest redshifts around $z\sim9$ 
and $\nssim2\times10^7$ cells at the lower
redshifts around $z\sim3$, where the shell is physically larger). 
Similar flux measurements were found
for shell thicknesses between 
$\nssim0.1\mbox{--}2.5\%$ of $r_{\mathrm{vir}}$, 
which implies that there is some flexibility in the value assigned to $\Delta r$.
A larger grid than $\mathcal{G}_{0.1}$ is therefore required, in order
to compensate for the half of the shell that spills
beyond the $0.1r_{\mathrm{vir}}$ region. The
filaments are thus constructed over a $0.15r_{\mathrm{vir}}$ region 
according to step $3$ of
the colour propagation scheme described in Appendix \ref{tracer_appendix}.
This particular grid length is chosen 
because the physical width of the cells within $\mathcal{G}_{0.15}$ is still
at the $12\,$pc limit for every time output, and so the extra grid length makes
no difference to the accuracy of the mass and angular momentum computations.

It is clear from the above that 
a decision has been made to represent the quantities $r_m$ and $\Delta r$ as fractions of the host's
virial radius, which grows with time in the simulation.
The spherical flux method hence sweeps up gas by construction,
and a simple trapezium integration of radial flux signals across the shell could miss this component
if the filament's trajectory is mostly azimuthal. 
We therefore recalculate $\Delta M_{\mathrm{flux}}$ (and $\Delta \bmath{J}_{\mathrm{flux}}$) to explicitly take 
the swept up component into account by: 
\begin{enumerate}
\item fixing the sphere radius $r(t_1)$ over the given time interval $\Delta t_{21}(\equiv t_2-t_1)$ between time 
outputs $t_1$ and $t_2$;
\item applying equations (\ref{eqn_am_flux}) and (\ref{eqn_mass_flux}) to these identical spheres at the two time outputs,
and performing the trapezium integration in equation (\ref{eqn_trap_int}), and;
\item combining the above spherical flux signals with the 
corresponding instantaneous mass ($\Delta M_{\mathrm{stat}}$) and angular momentum ($\Delta \bmath{J}_{\mathrm{stat}}$)
of all the filament cells measured at the later time output $t_2$ within the small spherical shell spanning the volume 
between $r(t_1)$ and $r(t_2)$.
\end{enumerate}
The integrated quantities at time $t$ hence correspond to:
\begin{eqnarray}
M_{\mathrm{T}}(t) &=& \sum_{t_i}^t [\Delta M_{\mathrm{flux}}(t^{\prime}_j) + \Delta M_{\mathrm{stat}}(t^{\prime}_j)]\label{eqn_comp_mass}\\
\bmath{J}_{\mathrm{T}}(t) &=& \sum_{t_i}^t [\Delta\bmath{J}_{\mathrm{flux}}(t^{\prime}_j) + \Delta\bmath{J}_{\mathrm{stat}}(t^{\prime}_j)],
\label{eqn_comp_angmom}
\end{eqnarray}
where the first and second terms refer to the signals measured using
the spherical flux method and the `static shell' approximation used for the swept up gas, respectively. 
The ratio of these terms in equation (\ref{eqn_comp_mass})
satisfies $\Delta M_{\mathrm{flux}}/\Delta M_{\mathrm{stat}}\gtrsim10$ for every time output in the simulation, 
indicating that the amount of gas swept up over a given time interval
is much smaller than the radial mass flux (which is somewhat expected if the output sampling is fine enough). 
The initial time output in equations (\ref{eqn_comp_mass}) and (\ref{eqn_comp_angmom}) is denoted by $t_i$
and the sum is performed over all discrete time outputs $t_i\leq t^{\prime}_j\leq t$,
yielding projections along the disk rotation axis:
\begin{equation}
J_{p}(t) = \bmath{J}_{\mathrm{T}}(t)\cdot\bmath{\hat{J}}_b(t).
\label{eqn_proj_filhot}
\end{equation}

\subsubsection{The satellite contributions to $m_b$ and $\bmath{J}_b$}\label{satellite_contributions_section}

If equation (\ref{eqn_mass_flux}) were applied to the satellites, it could
yield zero mass accretion events as 
satellite mergers are discrete burst events that can easily pass undetected 
through the shell over a given time interval.
The following simple prescription, which seeks to
pinpoint mergers with the disk, has been used to 
measure the instantaneous mass and angular momentum
of gas belonging to satellites before it is accreted onto the central galaxy
(satellite stars are ignored in this study as they are excluded from
the baryonic disk definition in equation \ref{eqn_disk_mass}).
The algorithm begins by finding all the 
luminous satellites (according to the
method described in Appendix \ref{satellite_appendix}) 
that infringe the $0.1r_{\mathrm{vir}}$ and $r_{\mathrm{vir}}$ regions at $t$
and $t+\Delta t$ respectively. Each satellite identification number 
at $t$ is then compared
with the list of satellite identifiers at $t+\Delta t$. Without a match, 
a given satellite has either fully merged with the disk
or has been stripped of mass upon crossing the central region and 
has consequently passed below the minimum
resolution limit of $100$ particles imposed by the clump-finder. 
The merger has hence already taken place or is due to take place
very shortly, and so the satellite signals at $t$ are measured.
If, however, the 
satellite is resolved at $t+\Delta t$ (it may have moved closer to the disk
or outside the central region during the merger process), 
then it has evidently not yet been accreted, and so no signal is recorded.
The gas within the satellite virial region is hence included in the 
cumulative signal at $t$ provided the satellite is:
\begin{enumerate}
\item encroaching the $0.1r_{\mathrm{vir}}$ region at $t$, and; 
\item unresolved at $t+\Delta t$.
\end{enumerate}
The satellite gas mass $M_s$ and angular momentum $\bmath{J}_s$ deposited within
$0.1r_{\mathrm{vir}}$ at $t$ are subsequently computed using direct
summation and equation (\ref{J_eqn}) respectively,
and are coupled with the totals from the previous
epochs in an analogous fashion to equations (\ref{eqn_comp_mass}) and (\ref{eqn_comp_angmom}):
\begin{eqnarray}
M^s_{\mathrm{T}}(\leq0.1r_{\mathrm{vir}}(t),t) &=& \sum_{t_i}^{t} M_s(\leq0.1r_{\mathrm{vir}}(t^{\prime}_j),t^{\prime}_j)\label{eqn_Msat}\\
\bmath{J}^s_{\mathrm{T}}(\leq0.1r_{\mathrm{vir}}(t),t) &=& \sum_{t_i}^{t} \bmath{J}_s(\leq0.1r_{\mathrm{vir}}(t^{\prime}_j),t^{\prime}_j),
\label{eqn_Jsat}
\end{eqnarray}
yielding a projected component along the disk rotation axis:
\begin{equation}
J^{s}_{p}(t) = \bmath{J}^s_{\mathrm{T}}(t)\cdot\bmath{\hat{J}}_b(t).
\label{eqn_proj_sat}
\end{equation}

\noindent The explicit time dependences of $J_p,\bmath{J}_{\mathrm{T}}$ and $\bmath{J}_b$ 
in equations (\ref{eqn_proj_filhot}) 
and (\ref{eqn_proj_sat}) are henceforth dropped for brevity.

\subsubsection{Assumptions}
By comparing  at each epoch the projected angular momentum of gas measured using equations (\ref{eqn_proj_filhot}) 
and (\ref{eqn_proj_sat}) with the disk signal found using equation (\ref{eqn_disk_am}), two 
implicit assumptions are being made:
\begin{itemize}
\item the angular momentum of the material passing
through the spherical shell at $0.1r_{\mathrm{vir}}$ as measured at $t$
reflects the amount deposited onto the disk at the time of contact $t+\delta t$
(i.e. $\delta t$ is assumed to be smaller than the time interval between two consecutive outputs of the simulation $\Delta t$), and;
\item the time integrated filament and hot gas
flux signals are well approximated by a trapezium integration. 
\end{itemize}

\noindent The first condition draws attention to possible time lags between 
material crossing the shell and
accreting onto the disk, and its
importance can be estimated by
comparing the free-fall time $t_{\mathrm{ff}}$ 
at $0.1r_{\mathrm{vir}}$ with the separation
between consecutive time outputs in the simulation $\Delta t$
(for a defense of the view that filaments are indeed 
well approximated by radial free-fall, see e.g. \citealt{Rosdahl12}). 
In the \nutco run it was found that the range in ratio of these two timescales
satisfied $0.1\lesssim t_{\mathrm{ff}}/\Delta t\lesssim1$, 
and so it is likely that the
change in system configuration during the interval does not 
yield a significantly different amount of angular momentum transferred 
to the disk compared with that measured
at $0.1r_{\mathrm{vir}}$, thereby justifying the first assumption.
The second assumption is considered in
the following section.

\section{Results}\label{results_section}

The techniques described in the 
previous section are now implemented in an attempt to examine whether: 
(i) filaments dominate $\bmath{J}_b$ at high redshift, and; 
(ii) there are any special episodes of disk angular momentum acquisition.

\begin{figure*}
\centering
\includegraphics[width=1.5\columnwidth,height=9.125cm]{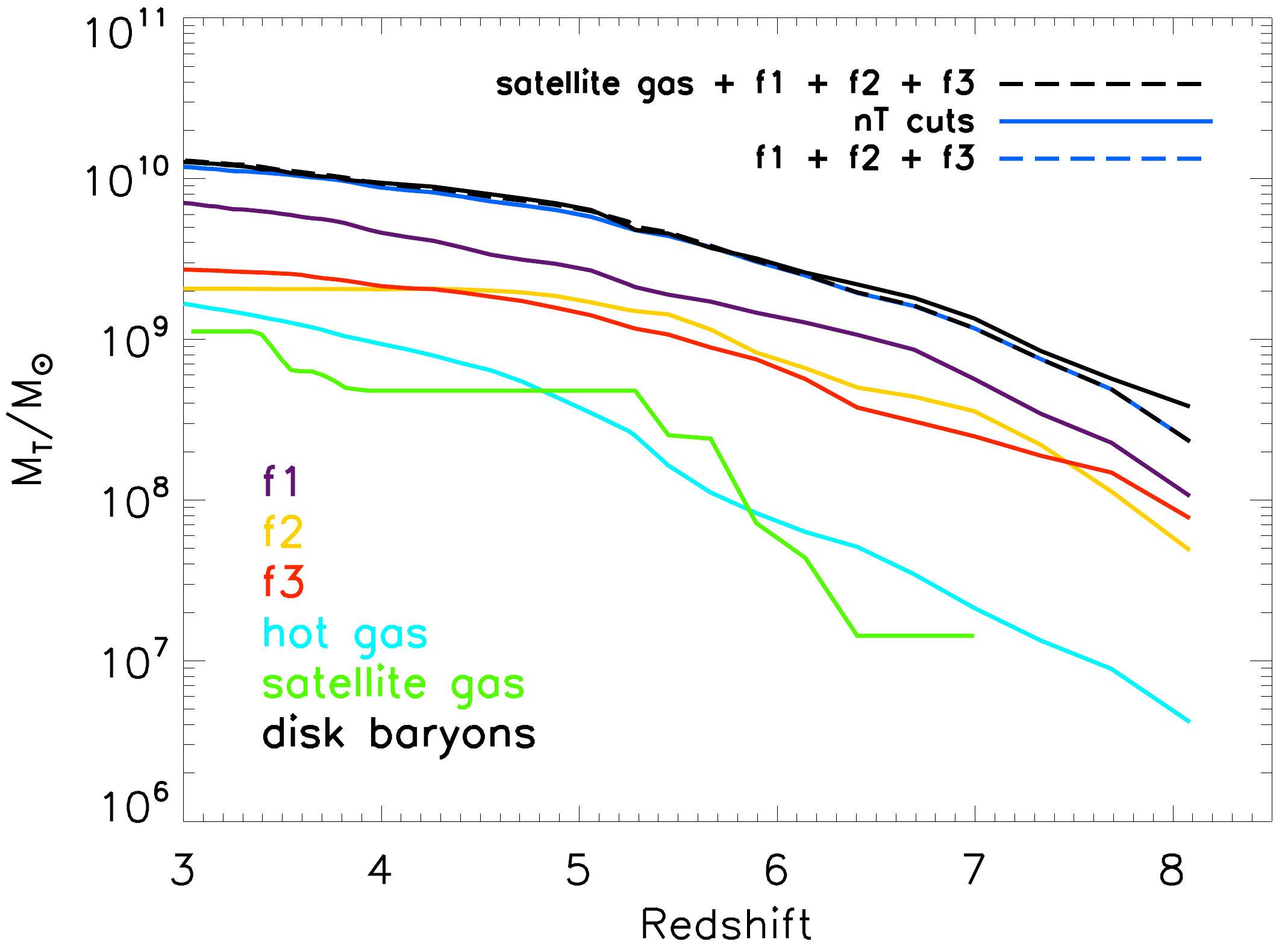}
\includegraphics[width=1.5\columnwidth,height=9.125cm]{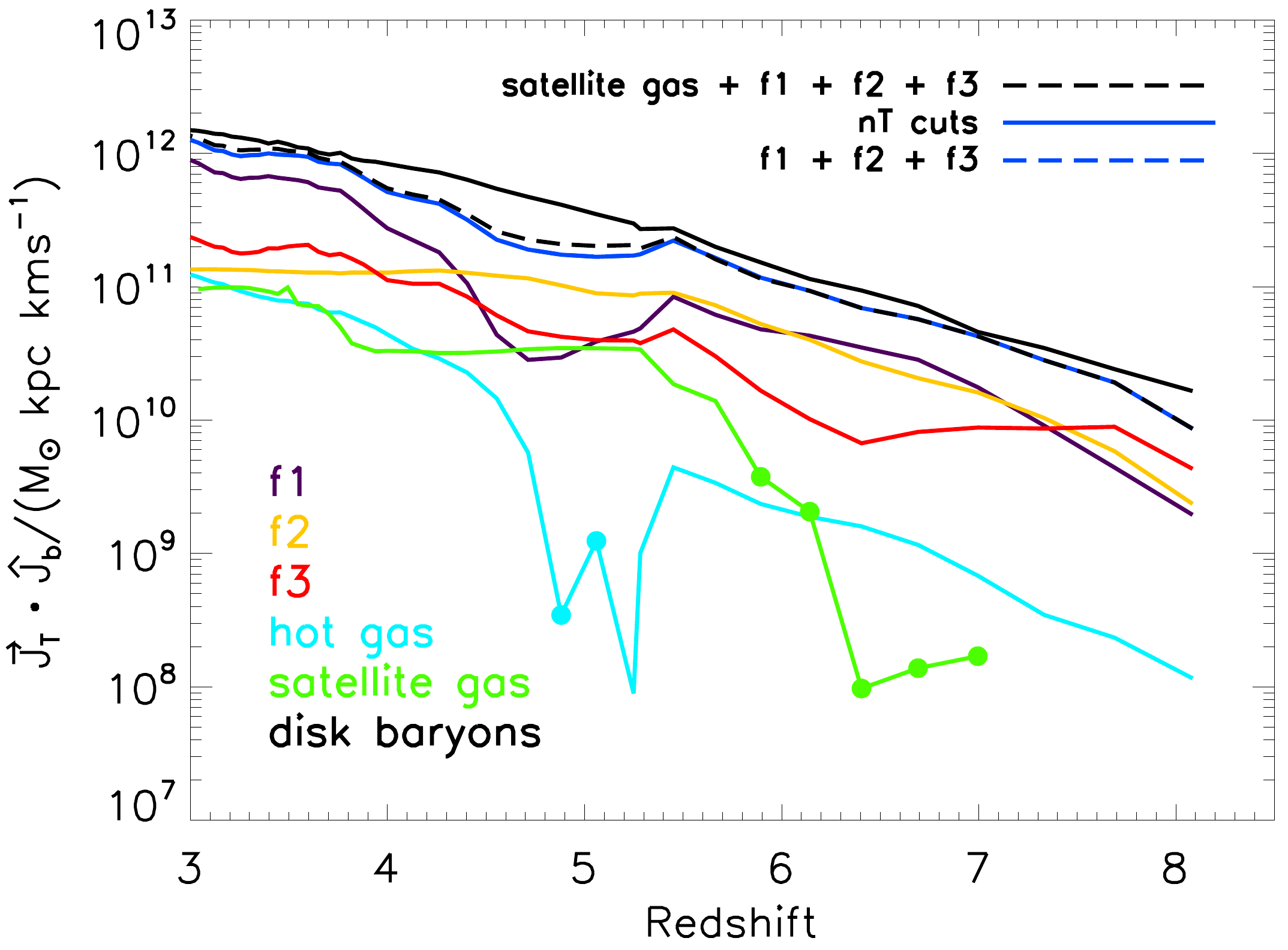}
\caption{The relative importance of hot gas ($T>2\times10^5\,$K), 
cold filament gas and satellite gas
in establishing the mass and angular momentum budget of the central
disk at high redshift. $M_{\mathrm{T}}$ (upper panel) and 
$\bmath{J}_{\mathrm{T}}$ (lower panel) 
have been accumulated with their corresponding signals from all prior epochs. 
The projection of $\bmath{J}_{\mathrm{T}}$ along the unit disk direction 
$\bmath{\hat{J}}_b$ then yields $J_p(z)$ in equation (\ref{eqn_proj_filhot})
for the filament and hot gas phases.
A similar computation has been performed for the satellites,
except that equations (\ref{eqn_Msat}) and (\ref{eqn_Jsat}) have been used in 
place of equations (\ref{eqn_comp_mass}) and (\ref{eqn_comp_angmom}), 
with the filled circles corresponding to negative $J_p(z)$ contributions.
The dashed dark blue lines sum over the individual filament signals
at each epoch, and cannot be distinguished from the solid dark blue lines,
which show the accumulated signals for all of the filament gas
found by applying the \nT cuts. The mass and angular momentum
of the disk baryons defined by equations (\ref{eqn_disk_mass}) and (\ref{eqn_disk_am}) 
have been computed at every time output and are given by the solid black lines.}
\label{projection_plots}
\end{figure*}

Fig.~\ref{projection_plots} plots the redshift evolution of
the integrated net inflow of mass towards the disk
in the upper panel, and the projection of 
the integrated angular momentum along the disk's axis of rotation 
in the lower panel. All of the measurements have been performed
at or within the $0.1r_{\mathrm{vir}}$ boundary as described in Section \ref{am_methods}.
Baryonic disk signals correspond to the solid black lines, 
while the satellite gas (equations \ref{eqn_Msat} and \ref{eqn_proj_sat})
and hot gas \mbox{(equations \ref{eqn_comp_mass} and \ref{eqn_proj_filhot})} signals
are given by the green and cyan lines respectively,
with the filled circles in the lower panel indicating negative cumulative angular momentum projections.
Each filament is distinguished by its unique colour,
and the dashed dark blue line represents their sum at 
each epoch \mbox{(equations \ref{eqn_comp_mass} 
and \ref{eqn_proj_filhot})}.
The solid dark blue line shows the signals of all of the filament cells
that have been found by applying the \nT criteria at each redshift, and 
serves as a consistency check for 
the individual filament computations. The 
dashed and solid dark blue lines are indistinguishable from one another in both panels, which implies
that the vast majority of the filaments cells have been accounted for,
and reinforces the high level of accuracy of the tracer colouring technique.
Finally, it is expected that the total gas angular momentum and mass contributions from 
the filament and satellite gas phases 
are equal to the baryonic disk signals at each epoch.
The dashed black lines in Fig.~\ref{projection_plots} test this hypothesis 
and should be compared with the corresponding solid black lines.

The disk, filament, hot gas and satellite signals at $z_{\mathrm{in}}^{\prime}\sim8.5$
are subtracted from their counterparts at $z<z_{\mathrm{in}}^{\prime}$ and are first recorded 
at $z\sim8$ in Fig.~\ref{projection_plots} for reasons associated with the tracer colouring algorithm
(which could not be applied until $z_{\mathrm{in}}\sim10$, the earliest
time output that displayed a smooth, distinct filament configuration in the \nutco run),
and the subsequent trapezium integration of equations (\ref{eqn_am_flux}) and (\ref{eqn_mass_flux}).
This approach of shifting the origin of the entire system forward to $z_{\mathrm{in}}^{\prime}$
and modelling the $0.1r_{\mathrm{vir}}$ sphere as being empty 
at $z_{\mathrm{in}}^{\prime}$ is justified given:
a) the desire to maintain consistency with the starting point of the tracer colouring algorithm, and;
b) the negligible amount of time elapsed between $z_1$ 
(the very first output from the simulation)
and $z_{\mathrm{in}}^{\prime}$
as compared to the time elapsed between $z_1$ and the final output at $z=3$. 

\begin{figure*}
\centering
\includegraphics[width=1.5\columnwidth,height=7cm]{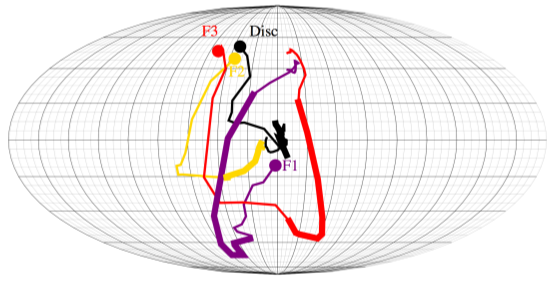}
\includegraphics[width=5.6cm]{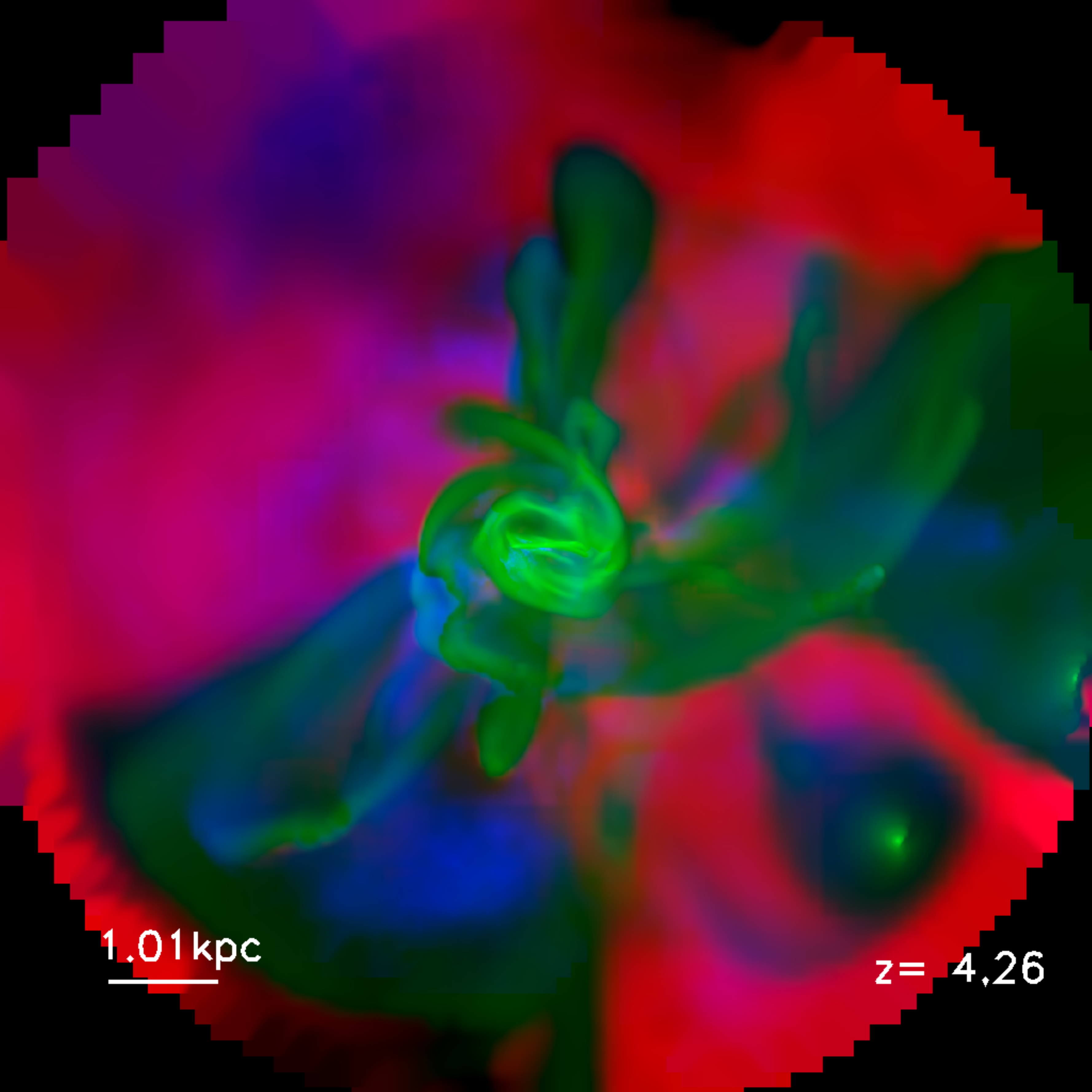}
\includegraphics[width=5.6cm]{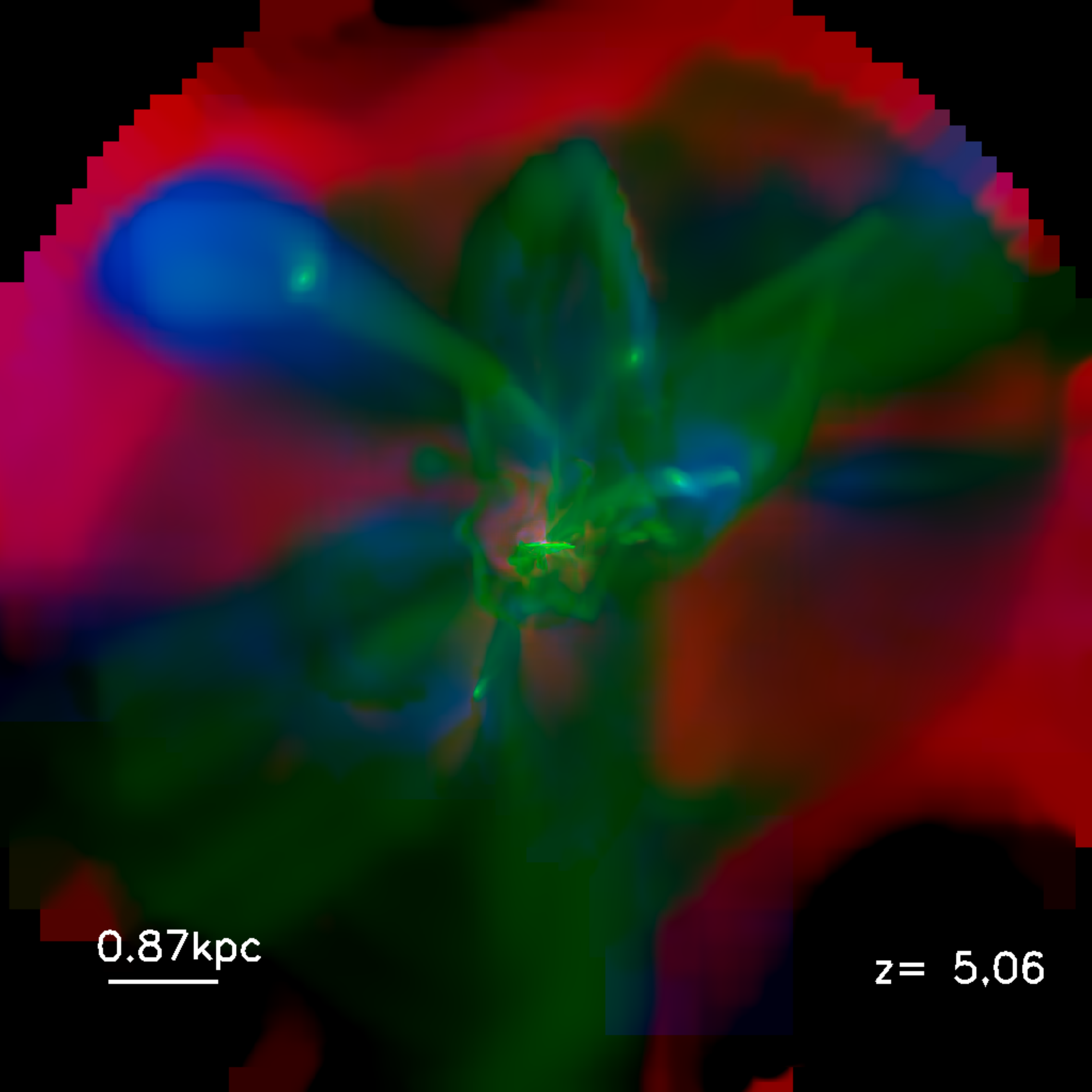}
\includegraphics[width=5.6cm]{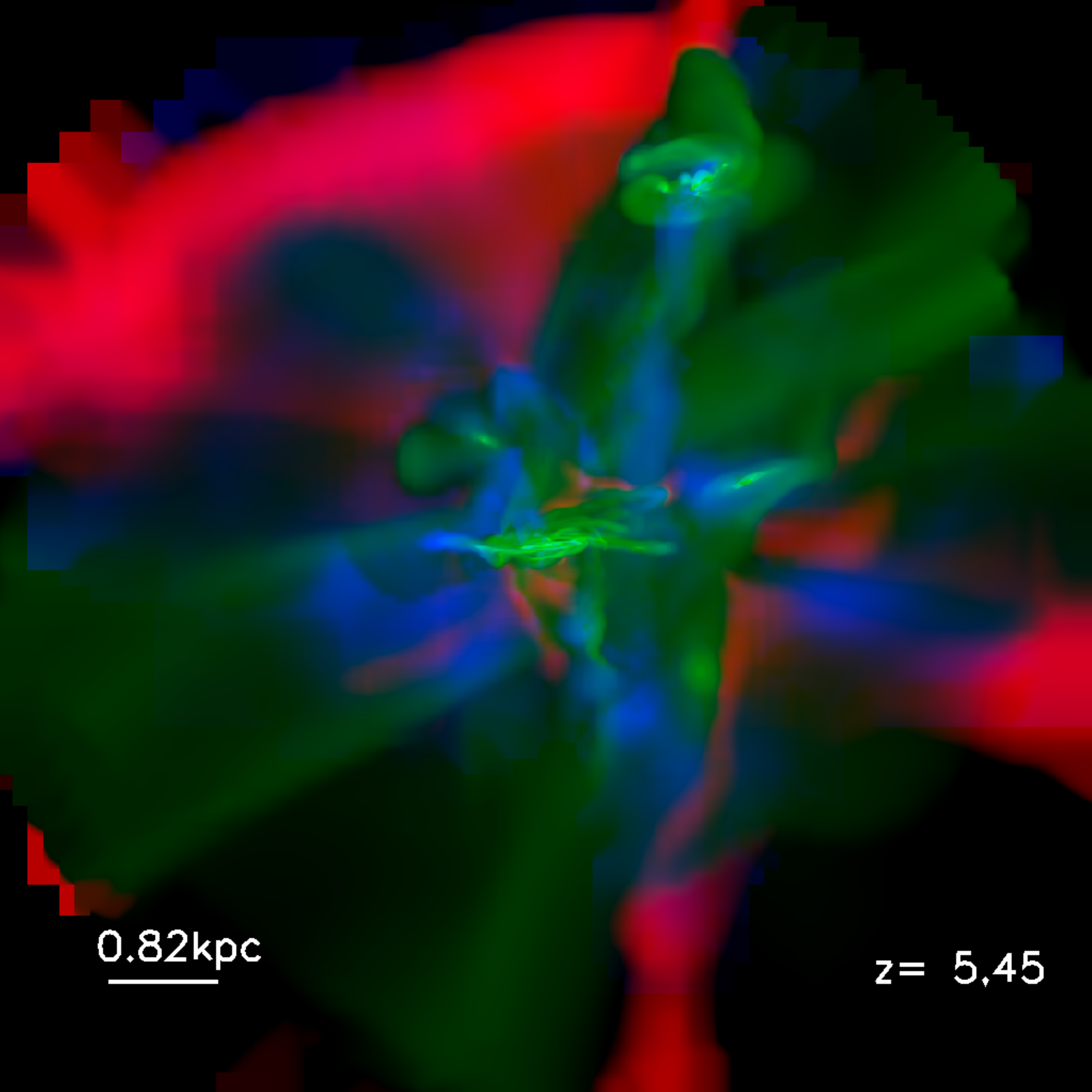}
\caption{Top panel: A Mollweide projection of the trajectories followed by
the cumulative angular momentum vectors of the disk and
filaments as a function of redshift, colour coded as in Fig.~\ref{projection_plots}. 
The filled circles indicate the starting point ($z=8$) and the thick parts 
of the curves mark the redshift interval $4.5 \lesssim z \lesssim 5.5$ where the cumulative angular momenta 
of the filaments undergo a dip in Fig.~\ref{projection_plots}. 
Bottom panels: RGB images showing the $z$-projected temperature (red channel),
density (green channel), and radial velocity (blue channel) of the gas enclosed 
within half the virial radius of the host halo during the redshift 
interval $4.5 \lesssim z \lesssim 5.5$. Redshifts and scales are indicated at the bottom of each panel. 
Note the extremely disordered gas density field at $z=5.06$ close to the (smaller) disk in the middle bottom panel, as compared 
to the previous (bottom right panel) and later (bottom left panel) 
epochs. This disorderliness is correlated with both the minimum of the dip in 
filament angular momentum shown in Fig.~\ref{projection_plots} and the 
dramatic change in the direction of the angular momenta of all the filaments visible in the top panel above.}
\label{dip_explanation}
\end{figure*}

\subsection{The mass growth of the central disk component}

It appears that the technical details surrounding the choice of starting point  
from the previous section are somewhat negligible because the upper panel of Fig.~\ref{projection_plots} indicates
that the baryonic disk experiences rapid growth 
at high redshift, where its mass increases by
a factor of $\nssim8$ between $z\sim8$ and $z\sim6$. 
This growth is slower at later times, with a factor
of $\nssim2$ increase in disk mass between $z\sim5$ and $z=3$. 
The periods of upturn in the green curves correspond to discrete luminous satellite merger events,
which deposit $\nssim1\%$ of the disk's mass in the form of gas at early times ($z\sim7$)
and $\nssim10\%$ at late times ($z=3$). Satellites do not therefore
appreciably affect the mass budget of the disk in the redshift range $3\lesssim z\lesssim8$. 
The amount of hot gas accreted onto the disk increases at a very stable rate
owing to the long cooling times associated with a metal-poor intergalactic medium
in the \nutco run, but by $z=3$, the hot phase only accounts for about
one tenth of the disk's accumulated mass. 
This phase thus plays 
a similarly subdominant role to the luminous satellites.
Evidently the mass budget of the disk is largely
controlled by the behaviour of filaments, and Fig.~\ref{projection_plots} suggests that it is the purple
filament corresponding to region $1$ in the left panel of Fig.~\ref{filaments_visualization} that 
carries the most mass to the disk, typically transporting at least 
twice as much mass at a given epoch compared with the yellow and red filaments, whose 
contributions flow in approximately equal measure until $z\sim5.5$ whereupon the purple and yellow
filaments merge, with the former surviving (Fig.~\ref{distr_tracer}). The disappearance 
of the yellow filament is evident from the plateau in its accumulated 
mass evolution for $z\lesssim5.5$.

By comparing the dashed black line with the solid black line in the upper panel of Fig.~\ref{projection_plots},
it can be seen that the spherical flux method (which dominates the signal in equation \ref{eqn_comp_mass}), designed
to predict the net radial inflow of gas accreted onto the disk, conserves
mass to high precision across the disk's evolution. 
This is remarkable given that the 
predicted flux rates across the shell could, in principle, vary quite dramatically
between time intervals and hence not be well approximated
by a trapezium integration. For example, it is possible that fast moving clumps
of material that are accreted onto the disk pass through the thin shell undetected,
which might lead one into thinking that the flux method has a systematic tendency
to underestimate the accumulated mass signal. The forepart of this argument appears
to hold for the very first recorded mass signal at $z\sim8$, where the
inflow of gas is just under a factor of $2$ below the disk signal in place
at that epoch. Clearly this is only a phenomenon that occurs at high redshift,
as it quickly becomes unimportant when the disk experiences rapid smooth growth 
via cold gas accretion over the subsequent time interval, which is captured by the flux method.
Even so, for completeness, it is interesting to speculate on this high redshift discrepancy.
One might propose that by increasing the thickness of the spherical 
shell from $\Delta r=1\%$ of $r_{\mathrm{vir}}$ to a larger
value, there is a higher probability of capturing this apparent `missing' filament mass. Yet 
it has been found that even when 
$\Delta r\sim10\%$ of $r_{\mathrm{vir}}$ is used at the first few time outputs,
there is a negligible difference in the filament (and hence filament plus satellite gas) signals.
The insensitivity of the flux rate to these changes at high redshift symbolizes
the robustness of the spherical flux method and
implies that the offset is not caused by an underestimation of the amount of mass transported
radially inward by the filaments.  
Given the large $100$ particle threshold imposed by the clump-finder and the technique
adopted by the satellite-finding algorithm in Section \ref{satellite_contributions_section} 
of identifying satellites within $0.1r_{\mathrm{vir}}$,
there does appear to be some scope for suggesting that the 
luminous satellite component is underestimated at this epoch: less massive, darting satellites 
could be excluded from the light green curves in Fig.~\ref{projection_plots}. 
After all, the satellite contributions may appear somewhat artificial in the sense that it is unlikely
that there are no satellite merger events beyond $z\sim7$.
We therefore surmise that this issue could be resolved
by including the contribution of satellite galaxies coming from further away than $0.1r_{\mathrm{vir}}$ at 
high redshift and possibly reducing the particle threshold of the clump-finder,  
although this latter option would impact the reliability of our satellite mass 
and angular momentum measurements.
In light of these difficulties, and the fact
that this single interval of seeming lack of mass conservation constitutes the least relevant 
high redshift interval in Fig.~\ref{projection_plots},
further analysis probing this discrepancy is deemed unnecessary. 
Indeed, such an underestimate will only have a minor effect for $z\lesssim7.5$ 
given the rapid rate of disk growth at early times.  

\subsection{Which components dominate the disk's angular momentum budget?}

The general trends of
satellites and hot gas contributing a weak mass signal
are also imprinted in their angular momentum evolution,
shown in the lower panel of Fig.~\ref{projection_plots}.
At $z=3$, the hot gas signal (like the satellite signal) accounts
for roughly one tenth of the disk's accumulated angular momentum, 
and with the exception of its evolution between $4.5\lesssim z\lesssim5.5$ 
(discussed in more detail below), increases at a steady, uniform rate towards late times.
In terms of angular momentum modulus, it therefore appears
that the relative contributions of the satellite and hot gas components 
to the disk angular momentum budget generally trace
their relative disk mass contributions. 
In terms of angular momentum direction, the satellite material which ends up deposited
onto the central disk has a tendency to be anti-aligned 
with respect to the disk's angular momentum direction at high redshift, but because the signal strength is relatively weak,
this mismatch only has a negligible effect. Certainly at redshifts
below $z\sim4.5$, the mass that is headed towards the disk in the satellite (and hot gas) phase 
appears to be more closely aligned with the disk's direction.

When integrated across all three components (solid and dashed dark blue lines),
the filament angular momentum signals show a similarly strong correlation 
with their corresponding mass trajectories. Clearly for the
redshift ranges $3\leq z\lesssim4$ and $5.5\lesssim z\lesssim7.5$, 
the disk's budget is driven by the transport of angular momentum from filaments.
When decomposed into its individual separate trajectories, however,
the cold phase shows noticeably different behaviour:  
(i) the yellow and red filaments transport
unequal amounts of angular momentum along the disk's direction 
for $z\gtrsim4.5$, in contrast with their relative mass contributions, and; 
(ii) the purple and yellow filaments are equally dominant for $z\gtrsim5.5$,
unlike in the mass plot where the purple filament clearly dominates at all epochs. 
It should also be noted that the absence of a correlation between negative satellite signals
and negative filament signals
implies that satellites are not necessarily sensitive
to the dynamics of the filament flow, despite often
streaming along these flows (e.g. Fig.~\ref{satellites_im}).

Arguably the most interesting feature of Fig.~\ref{projection_plots}
is the lack of response the disk has to the sudden dip in the projected component of the
purple filament's cumulative angular momentum during the 
filament merger phase between $4.5\lesssim z\lesssim5.5$.
It was found that the change in the direction of the disk's axis of rotation
during successive intervals across this period was negligible (of the order a few degrees, 
as can be seen in the top panel of Fig.~\ref{dip_explanation}), 
and so the sharp drop-off implies that the purple filament transports
angular momentum that is anti-aligned with the disk
during the early phase of the merger (equation \ref{eqn_proj_filhot}).
The red filament's disk-projected angular momentum contribution also abruptly changes
around $z\sim5.5$, which
when combined with the purple filament's, yields
the flattened integrated dark blue curve that reflects the net small amounts of angular
momentum being transported inwards by the filaments during this merger period.
Looking at the bottom right and middle panels of Fig.~\ref{dip_explanation}, it becomes clear 
that this disruption is caused by the ongoing merger  of the central disk with a quite massive 
satellite galaxy. This process leads to a complete reorganisation of the gas density field in the central region where 
the filaments connect to the disk, as shown in the bottom middle panel of Fig.~\ref{dip_explanation}.
Whilst this does not affect the mass flow, as the gas must eventually pile up onto the 
disk in the absence of feedback and shock heating, it has dramatic 
consequences for the angular momentum of the gas embedded in the filaments. 

One may therefore expect the new population of disk stars
that form from this anti-aligned angular momentum filament material 
to slightly reduce the disk angular momentum growth rate that is observed in the lower panel of Fig.~\ref{projection_plots}.
However, a new `equilibrium' configuration between filaments and disk rapidly arises, presumably 
through torquing (see bottom left panel of Fig.~\ref{dip_explanation}), in which  
the purple filament's angular momentum contribution along the disk's direction is restored. 
Moreover, at $z\lesssim4$, the purple filament becomes the 
dominant filament contributor to the angular momentum budget
by at least a factor of $\nssim3$. This prominence is partly fuelled 
by the large-scale merger between the purple and yellow filaments 
that begins at $z\sim5.5$ and induces a change in the mass, velocity and trajectory 
direction of the gas embedded within the surviving filament.
Note that by $z=3$, which corresponds to the end of the simulation, the angular momentum 
directions of the gas present in the red and purple filaments are fairly well aligned with one another
but not with the disk (see the top panel of Fig.~\ref{dip_explanation}), which we interpret as small-scale torquing 
of large-scale driven quantities.  

Meanwhile, the angular momentum that our spherical
flux method measures from filaments at $0.1r_{\mathrm{vir}}$ 
differs from the filament angular momentum
deposited onto the disk during the satellite merger period
because the inner dynamics within $0.1r_{\mathrm{vir}}$
are severely perturbed (note that the resultant complex 
inner gas trajectories appear to be imprinted in the hot gas
signal too, which becomes anti-aligned in Fig.~\ref{projection_plots} around $z\sim5$). 
It is likely that isothermal shocks in the disk boundary region are 
responsible for filtering aligned angular momentum to the disk during this phase, 
driving its angular momentum growth, because the baryonic disk angular momentum is dominated 
by its stellar component ($\abs{\bmath{J}_{\star}}/\abs{\bmath{J}_{\mathrm{gas}}}\sim2.6$ at $z\sim5$ for example),
which seems to stabilise it against strong local perturbations. Should this effect apply 
to a large fraction of disk galaxies, one could conclude  
that contrary to dark matter haloes \citep[e.g.][]{Bett_spinflip_12}, 
minor mergers of galaxies do not seem capable of significantly 
altering the direction of the angular momentum of the central disk. 

\section{Discussion}\label{discussion_section}

Perhaps the most defining feature of disks in galaxies 
with $M_{\star}\lesssim 3\times10^{10}\,\mathrm{M}_{\sun}$ is the amount of 
angular momentum they possess, as it is this property that establishes many
of their fundamental scaling relations such as the Tully--Fisher relation (e.g. \citealt{Governato07}) and 
those linking velocity, metallicity, surface brightness
and stellar mass \citep{Dekel_Woo_03,Dekel06}. 
The Tully--Fisher relation, in particular, is an important
constraint that many semi-analytic models of galaxy formation
strive to satisfy \citep{Hatton03,Cattaneo06,Croton06,DeLucia06}.
Understanding how the disk acquires its 
angular momentum is hence of central importance,
and has been the subject of this paper.

\subsection{The dominance of filaments at high redshift}

The evidence favouring the growth of disks in low mass systems via a
mode of cold gas accretion appears to be mounting. 
\cite{Keres05} performed SPH simulations of
several hundred galaxies between $0\leq z\leq3$ using
a comoving gravitational softening scale of $5\,h^{-1}\,$kpc,
and found evidence of a clear shift in the total fraction
of gas accreted with temperatures $T<2.5\times10^{5}\,$K
around a galaxy stellar mass of $M_{\star}\sim2.5\times10^{10}\,\mathrm{M_{\sun}}$, 
with at most a factor of $\nssim1.6$ deviation in $M_{\star}$
across this redshift range.
\cite{Ocvirk08} conducted a similar statistical analysis
on haloes from the \hmn simulation with 
$10^{10}<M_{\mathrm{H}}/\mathrm{M}_{\sun}<10^{13}$ 
at higher redshifts between $2\leq z\leq5$,
in an attempt to measure the temperature and density of the gas
deposited onto the central galaxies of these systems, which were simulated using the \ram code
at a physical resolution scale of $\sim 1 h^{-1} \,$kpc. Their results demonstrated that
the fraction of gas accreted at temperatures
$T<2.5\times10^{5}\,$K sharply increases 
for $M_{\mathrm{H}}\lesssim4\times10^{11}\,\mathrm{M}_{\sun}$,
in concordance with the estimates provided by
\cite{Birnboim03} and \cite{Keres05}.
By resolving and examining the various components of the filamentary gas phase
across a wide range in redshift,
the results from Fig.~\ref{projection_plots} in this paper provide quantitive support for
the cold mode paradigm of gas accretion onto a Milky 
Way like disk, advocated by the above studies. They also show that a
single filament can be responsible for driving the mass budget of the baryonic disk, 
although this does not necessarily map to a dominance of the disk's angular momentum budget,
as two filaments appear to closely share priority at the higher redshifts ($z\gtrsim5.5$), despite
a clear difference in their mass contributions.
The flow of cold gas onto the central disk is not necessarily a smooth, continuous process either.
While satellites do not significantly perturb the filament trajectories in general (except in the region close to the
central disk, and even then, not very often and only over short periods of time),
large-scale motions of filaments can lead to mergers between these cold streams \citep{Pichon11}
and this process is able to change the orientation of angular momentum advected along these flows,
probably even if gas channelled along each of the filament mergers in question is 
co-planar with the disk's rotation before the merger.
 
A quantitative result of this kind has remained elusive to many previous studies
mostly due to the parsec-scale physical resolution required to resolve the satellites, 
filaments and disk components
within the inner tenth of the virial region at $z>3$. Some authors
have recently started to probe this resolution barrier, however.
\cite{Kimm11} measured the angular momentum of gas in radial
bins as a function of redshift, for the same 
resimulated \nut Milky Way like halo analyzed in this paper. 
Upon stacking the radial profiles in two separate redshift
regimes ($z\leq3$ and $z>3$), they found evidence for a sudden loss
in the amount of specific angular momentum transported
by gas in the $0.1r_{\mathrm{vir}}$ region, but with a physical
resolution scale of only $\nssim50\,$pc, were unable to speculate
on the cause of this loss. Although the analysis in this paper does not address
this issue \emph{per se}, it does show
that the filaments supply enough angular momentum to the boundary of the disk region
to be able to account for the angular momentum locked-up in the disk across most epochs at $z>3$, a
result that has been found by using a 
filament tracer colouring technique. 

Several studies have also demonstrated that the angular momentum vectors of 
dark matter and gas are not necessarily aligned within the virial region of galaxies at
both low ($z=0$) and high ($z\leq3$) redshift \citep{Bett10,Roskar10}.
\cite{Danovich12} reported a weak correlation
between the angular momentum direction of gas in the inner ($\nssim0.1r_{\mathrm{vir}}$)
and outer ($\nssim1r_{\mathrm{vir}}$) galaxy regions at $z=2.5$,
and a large body of evidence has now been presented that confirms 
the existence of mismatches in gas angular momentum directions over
scales of a few disk scalelengths, giving rise to the population
of `warped' disks \citep{Garcia-Ruiz_02,Shen_Sellwood_06}. 
Despite being observed as a low redshift phenomena, the recent simulation study
by \cite{Roskar10} has argued that warps may exist around $z\sim2\mbox{--}3$.
In Fig.~\ref{projection_plots} of this paper,
the amount of angular momentum transported along the disk's direction
by all the filaments  is 
comparable to that locked-up in the disk,
which would not be the case if large-scale misaligned signals
reported above were preserved at the disk boundary. If the claim that
the directions of the gas angular momentum vectors
vary with distance from the central galaxy holds across the galaxy's lifetime,
some physical mechanism must be responsible
for aligning the infalling angular momentum at the disk's edge.
For example, \cite{Roskar10} have argued that freshly accreted gas at the virial radius is strongly
torqued by the hot halo gas component, and cited this as a possible cause of the 
warps between inner and outer disk structure. Whilst hot gas torques are unlikely 
to be dominant in our case, it is nevertheless surmised that
shocks and/or torques remove the components 
of the purple filament's velocity that are misaligned with respect to the disk's
plane of rotation, thereby explaining why this filament 
is able to maintain a relatively stable transfer of aligned angular momentum to the disk between $3\lesssim z\lesssim 8$.

Despite the evidence from both Fig.~\ref{projection_plots} and the aforementioned
simulation studies favouring the cold gas paradigm,
not all authors are convinced that this phase is quite so dominant
in growing the disk components of low mass galaxies. \cite{Murante12}
monitored the fractional accretion rate onto two Milky Way like haloes
and found that $\nssim50\%$ of the accreted gas onto the 
central galaxy between $3\leq z\leq6$ 
was in a warm phase, with a temperature in the range
$2.5\times10^{5}<T/\mathrm{K}<10^{6}$. They attributed this apparent discrepancy
to a supernovae feedback prescription that only modelled 
thermal heating, as opposed to thermal and kinetic heating. 
The effects of supernovae feedback on filament
structure were also examined by \cite{vandevoort11}, who suggested that
it reduces the growth rate of low mass central galaxies
residing in haloes with mass $M_{\mathrm{H}}\lesssim10^{12}\,\mathrm{M}_{\sun}$, 
implying that the filaments streaming
cold gas at large inflow rates in Fig.~\ref{projection_plots} do not survive when supernovae feedback is included.
\cite{Powell11}, however, argued that this result is probably an artefact 
of supernovae feedback implementation, because
with individual Sedov--Taylor blasts from the ultra-high resolution ($\Delta x_{\mathrm{res}}\sim0.5\,$pc) 
\nut feedback run simulation,
the net mass inflow rates of gas in the filament phase were found to be an order of
magnitude larger than the supernovae-driven mass outflow rates.
The \cite{Powell11} study hence implies that whilst
the spatial distribution of filaments local
to the disk is likely to be perturbed around sites of intensive
stellar explosions,
the amount of filament gas flowing towards the disk should largely be
unaltered. Therefore, the amount of angular momentum aligned with the disk that is transported by filaments
is probably unchanged too, given the general correlation
between the filament mass and disk-projected angular momentum signals in Fig.~\ref{projection_plots}.
It is therefore expected that the difference between the supernovae feedback
version of Fig.~\ref{projection_plots} and its
counterpart presented in this paper is not significant, 
but we leave this comparison as a future exercise.

\subsection{Understanding the impact of mergers}

Fig.~\ref{projection_plots} shows that luminous satellite mergers
constitute only a minor fraction of the disk's mass and angular momentum modulus between $3\lesssim z\lesssim8$. 
This result does not necessarily imply, however, that satellites as a general population have little
effect on the disk's evolution.
\cite{Bett_spinflip_12} analyzed present day Milky Way haloes of 
mass $M_{\mathrm{H}}\sim10^{12}\mbox{--}10^{12.5}\,h^{-1}\,\mathrm{M}_{\sun}$ from one
of the Millennium Simulation runs and examined
the importance of `spin flips', which are defined
as abrupt changes of more than $45^{\circ}$ in the orientation of a component's angular momentum vector.
They argued that over $90\%$ of their detected host halo spin flips were
caused by minor mergers, and demonstrated that the number of flip events
increases in the inner halo where the central galaxy resides. 
They further speculated that these spin flips
could destroy the host's stellar disk (which 
is included in the baryonic disk signal of Fig.~\ref{projection_plots}), 
or torque it (see, for example, \citealt{Ostriker89}). 
In both of these scenarios, one would expect a change in the angular momentum
direction of the baryonic disk component. The validity of this hypothesis remains an open question because
the existence of a simple correlation between dark halo
spin flips and disk spin flips is yet to be confirmed:
several recent studies have in fact hinted at a lack of correlation 
between the two \citep{Scannapieco09,Stinson10,Sales11}.
It will be interesting to test in more detail whether this claim by \cite{Bett_spinflip_12} holds for
the \nut host halo in this study, which experiences multiple minor mergers across its accretion history
(as indicated by Fig.~\ref{satellites_im}). However, we suggest that it is highly unlikely since the top panel of Fig.~\ref{dip_explanation} 
clearly shows that the angular momentum direction of the disk varies very little throughout the majority of its lifetime.

\subsection{High redshift contributions to the present day mass and angular momentum of the disk from the \nutco run}

It is also informative to estimate the 
fraction of the disk's mass and angular momentum at $z=0$ that
is already in place at $z=3$, as large high redshift contributions would further
highlight the importance of understanding the primordial
phase of Milky Way like galaxy growth.
Crude estimates are hence provided in this section, but
it should be noted that more rigorous measurements could
be made by analyzing all of the 
lower resolution outputs from the \nutco run between $0\leq z\leq3$
and extending Fig.~\ref{projection_plots} to $z=0$.

\cite{Kimm11} measured the total amount of baryonic mass (gas plus stars)
within $0.1r_{\mathrm{vir}}$ from the $\Delta x_{\mathrm{min}}=48\,$pc \nutco run at $z=0$ and found that
$M_b(z=0)\sim8\times10^{10}\,\mathrm{M}_{\sun}$.
This measurement includes the central contributions from all of the gas phases (satellite, hot mode, cold mode and disk)
and all of the stars (satellite, disk and bulge), and
is a factor of $\nssim6$ greater than
the baryonic disk signal at $z=3$ from Fig.~\ref{projection_plots}, which is
$M_b(z=3)\sim1.3\times10^{10}\,\mathrm{M}_{\sun}$. 
Multiplying the modulus of the total
baryonic specific angular momentum contributions within the $0.1r_{\mathrm{vir}}$ region
at $z=0$ from the \cite{Kimm11} study by the relevant component masses yields
an estimate of the total
baryonic angular momentum signal:
$J_b(z=0)\sim4.4\times10^{13}\,\mathrm{M}_{\sun}\,\mathrm{kpc}\,\mathrm{km\,s^{-1}}$.
This is a factor of $\nssim30$ higher than the modulus of the baryonic disk angular momentum
signal at $z=3$ shown in Fig.~\ref{projection_plots}:
$J_b(z=3)\sim1.6\times10^{12}\,\mathrm{M}_{\sun}\,\mathrm{kpc}\,\mathrm{km\,s^{-1}}$.
Note that the mass (and most probably the angular momentum) ratios
are upper limit estimates because the \cite{Kimm11} measurements include 
the satellite star and non-disk gas contributions, whereas Fig.~\ref{projection_plots}
just shows the baryonic disk signals defined according to 
equations (\ref{eqn_disk_mass}) and (\ref{eqn_disk_am}). 

In summary, at least $\nssim16\%$ ($\nssim4\%$)
of the \nut disk's final mass (angular momentum) is already in place by $z=3$, 
suggesting that high redshift epochs studied in this paper represent a
non-negligible period of the disk's accretion history.

\section{Conclusions}\label{conclusions_section}
We have analyzed an Adaptive Mesh Refinement cosmological resimulation
with $12\,$pc resolution in order to address whether the angular momentum acquired by
the baryonic disk of a Milky Way like galaxy at $z\geq3$
is driven by filaments. The filaments have been
identified by tagging Lagrangian tracer particles 
within a region of length $2r_{\mathrm{vir}}$ from the disk's centre
at each epoch. The method assigns one of three colours (one for each filament) to the 
tracer particles by using a nearest neighbours scheme,
and subsequently locates the `pure' cells within 
the spherical grid of radius $2r_{\mathrm{vir}}$ that contain tracer particles
of identical colour. Individual filament trajectories are then grown around these pure sites.
A satellite finding algorithm 
has also been presented, which operates
under the principle that a given satellite substructure of the main host at time $t$
has at some earlier time been a separate field galaxy of its own.
With the filaments and satellites resolved, the aim
of this paper has been to measure their 
contributions to the host's
disk angular momentum budget at each epoch.
For the filament and hot gas components, the mass and angular momentum
transported to the central disk has been computed
by recording both the net inward flux across an outwardly
propagating spherical shell on the disk's edge at $0.1r_{\mathrm{vir}}$, and the contribution from
the material swept up by the shell's outward radial motion. 
Satellite signals have been measured by adopting 
an alternative technique that pinpoints mergers with the disk
and records their angular momentum at the final pre-accretion stage.
The results suggest that:
\begin{itemize}
\item The cumulative effect of the cold filament gas phase
dominates the mass and angular momentum 
budgets of the disk as a function of time,
hence providing quantitative support
for the cold gas paradigm of disk growth in low mass
galaxies at high redshift.
\item For $5.5\lesssim z\lesssim7.5$, the largest portion of the filament
angular momentum signal is transported by two filaments,
which start to merge around $z\sim5.5$. This is accompanied by the central disk 
merging with a satellite galaxy, which temporarily induces a dramatic change in the orientation
of the angular momentum transported towards the disk,
but the surviving filament remnant quickly realigns itself along
the direction imposed by the large-scale flow and by $z\sim4$ dominates
the filament angular momentum budget by at least a factor of $3$.
\item The luminous satellites 
account for at most one tenth of the 
disk mass and angular momentum modulus at any given epoch.
\end{itemize}

\section*{Acknowledgements}
We thank Yohan Dubois for providing the tracer particle
data, and Taysun Kimm for useful discussions.
The \nut simulations were performed on the DiRAC
facility jointly funded by the Science and Technology Facilities Council (STFC), 
the Large Facilities Capital Fund of the Department for Business, Innovation \& Skills (BIS), 
and the University of Oxford.
The research of AS and JD is supported by Adrian Beecroft,
 the Oxford Martin School and the STFC.
HT is a grateful recipient of an STFC studentship.

\def\refname{REFERENCES}
\bibliographystyle{mn2e}
\bibliography{disk}


\appendix

\section{Identifying the filaments using tracer particles}\label{tracer_appendix}

\subsection{The algorithm}

The tracer propagation method, first encountered in Section \ref{tracer_section},
consists of three principal stages:

\begin{enumerate}
\item \textbf{Colouring the tracer particles across time.} At some initial early time $t_1$ ($z_1\sim10$),
planar cuts are used to divide the filaments 
into distinct regions within a spherical grid $\mathcal{G}_{2.0}$ of radius $2r_{\mathrm{vir}}$
(black planes in the left panel of Fig.~\ref{filaments_visualization}). 
Each region is assigned a unique colour and tracer particles 
are initially coloured according to the region in which they reside.
Hence there is a single filament per region
by construction, and we refer to the filaments in regions $1, 2$,
and $3$ of the left panel of Fig.~\ref{filaments_visualization} as the purple, yellow and red filaments respectively.

The Lagrangian nature of the system poses two computational 
difficulties when assigning colour:
\begin{itemize}
\item filaments mix in the vicinity of the disk
(as can be seen in the right panel of Fig.~\ref{filaments_visualization}), and;
\item filaments acquire a drift velocity on scales $r\gtrsim r_{\mathrm{vir}}$ \citep{Pichon11},
which can lead to them switching between regions within $r_{\mathrm{vir}}$ as the host grows.
\end{itemize}

\begin{figure*}
\centering
\includegraphics[width=1.61\columnwidth,height=5.1cm]{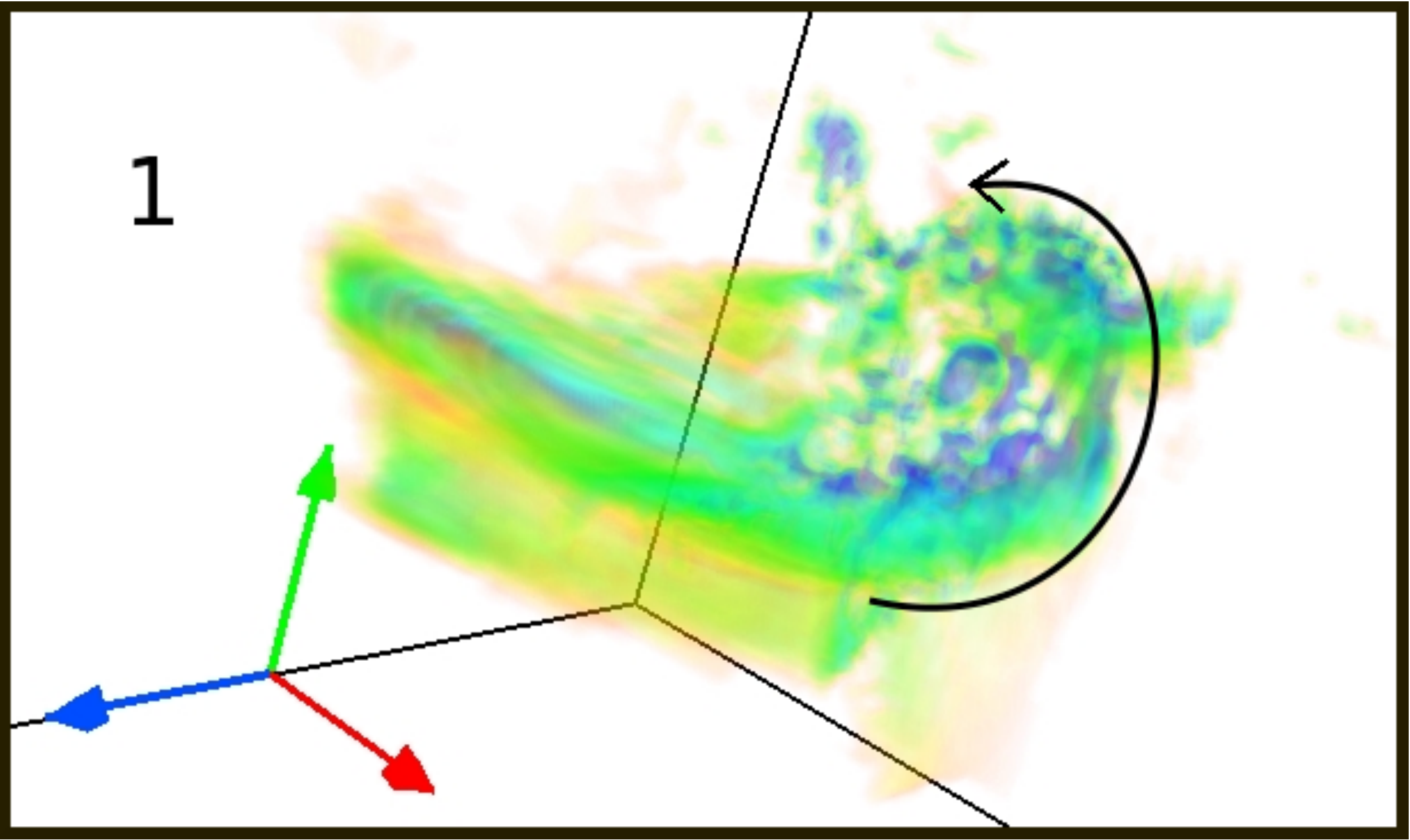}
\includegraphics[width=0.8\columnwidth,height=7.1cm]{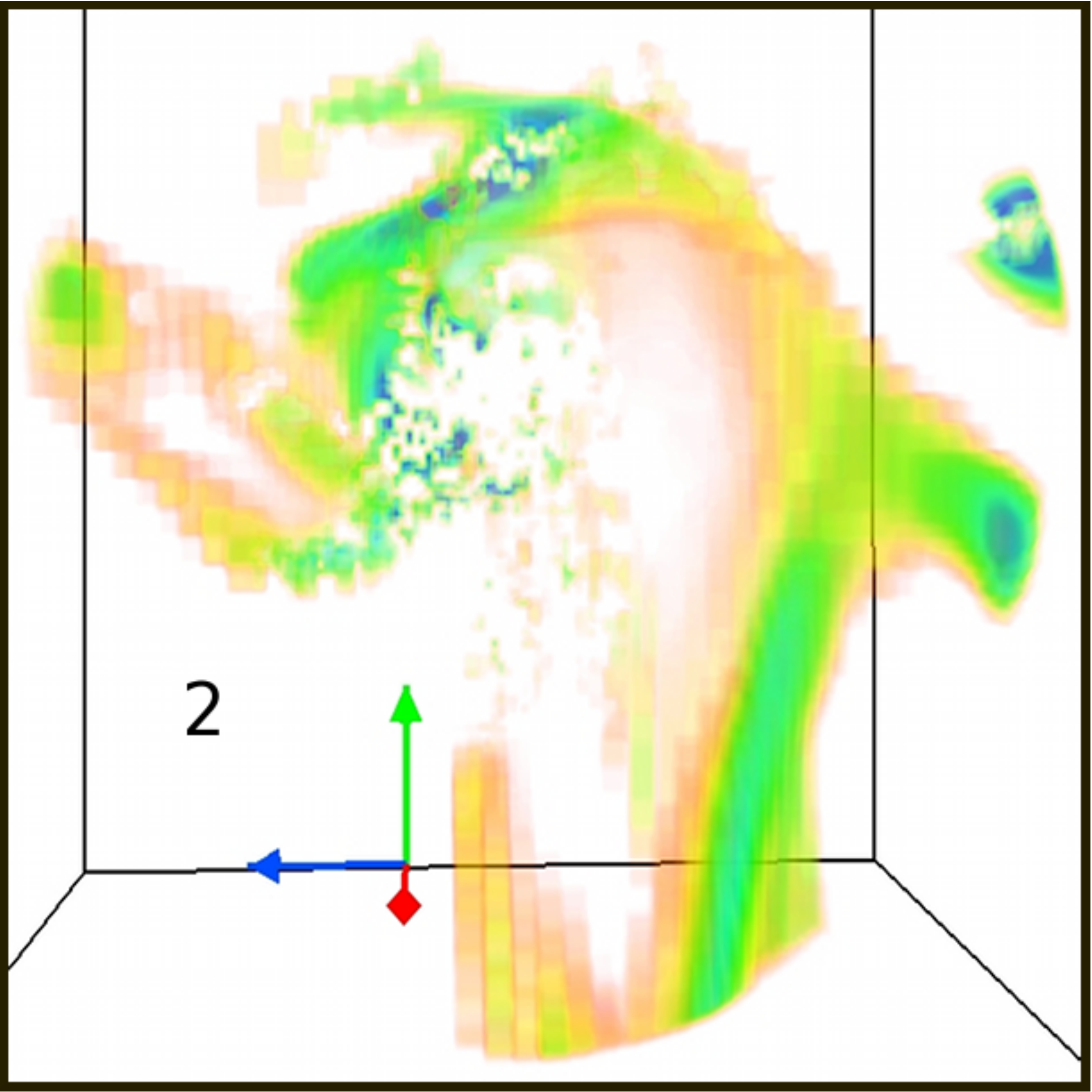}
\includegraphics[width=0.8\columnwidth,height=7.1cm]{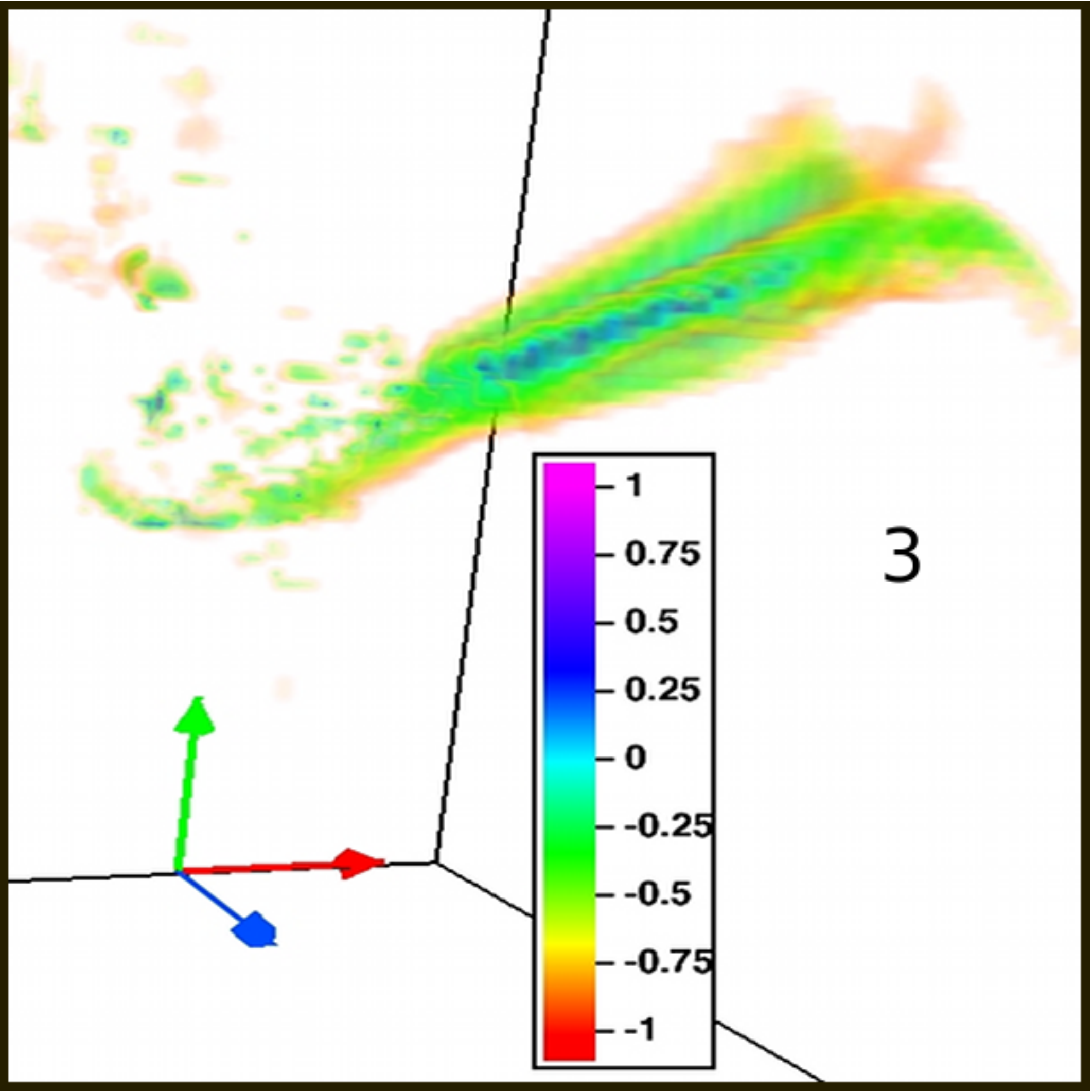}
\caption{These \vap images illustrate the trajectories of each
gas filament on $\mathcal{G}_{0.5}$ \mbox{($r_{\mathrm{vir}}\sim7.5\,$kpc)} at $z\sim8$ in the \nutco run with $\ell=17$ ($\Delta x\sim12\,$pc), 
as identified by the colour propagation technique. Each panel corresponds to the updated versions
of the purple, yellow and red filaments associated with regions $1, 2$ and $3$
respectively in the left panel of Fig.~\ref{filaments_visualization}, and follows the same
density colour scaling as in previous figures. The black curved arrow in the top panel indicates the gas flow direction of
the purple filament.}
\label{individual_filaments}
\end{figure*}

\noindent The first issue is tackled by resolving the filament trajectories at the following
time output $t_2$, by which time the vast majority
of the incorrectly coloured particles within $0.1r_{\mathrm{vir}}$ at
$t_1$ have accreted onto the disk during
the $\Delta t = t_2 - t_1$ interval,
and are therefore no longer 
associated with a filament.
To circumvent the second issue, 
previously flagged tracer particles $i$ at locations $\bmath{r}_{i}(t_1)$ 
that have moved during the $\Delta t$ interval
are updated according to their new instantaneous locations $\bmath{r}_{i}(t_2)$ at the following time $t_2$, 
provided that they coincide with filament cells at $t_2$.
All of the newly accreted tracer particles that are co-spatial with filaments 
within the (larger) host spherical region at $t_2$ that have not been
assigned a colour are then given a colour according to their host region.
At subsequent time outputs $t_j$ ($t_3$ onwards), 
the newly accreted tracer particles are instead assigned the majority
colour of their $N$ nearest coloured tracer
neighbours that sample the filaments at $\bmath{r}_i(t_j)$ ($N$ was at least $100$). 
Note that the above neighbouring scheme cannot be used for $t_2$ because
the constructed filament trajectories are not known until the end of the time output (see steps $2$ and $3$ below).

\item \textbf{Colouring the grid.} 
Once all of the filament tracer particles within $\mathcal{G}_{2.0}$ 
are coloured, the overall colours of the member tracer cells
are determined (particles located in non-filament cells within $r_{\mathrm{vir}}$ are ignored,
so that only filaments are resolved). 
The prescription used in this study enforces the condition that each coloured
tracer cell is `pure', consisting entirely of particles with identical colour.
The advantage of this approach
is that it avoids the use of \emph{ad hoc} criteria to help
decide the overall colour of a cell that contains
a certain ratio of particles of different colours.

\item \textbf{Individual filament growth.} 
The final step propagates the pure cell colours
to neighbouring uncoloured filament cells within $\mathcal{G}_{2.0}$. 
Filaments are simultaneously grown
around the cells with pure colour and the procedure terminates
once all neighbours of neighbours etc, are coloured and the only 
remaining neighbouring cells are non-filament cells or cells that already have a colour.
\end{enumerate}

\subsection{Uncertainties}
While the \nT cuts are only approximate, \cite{Powell11}
have argued that the associated uncertainties in the bulk physical properties 
of the filaments (average mass, velocity etc) remain very small
because: (i) these cuts successfully capture the dense core of the filaments 
(as confirmed by visual verification) around the disk
in the \nut suite runs with $\Delta x_{\mathrm{res}}=0.5\,$pc, and; (ii) reducing 
the lower density threshold has little impact on these properties.
Hence we also ignore the uncertainties associated with the \nT criteria in what follows,
and the constructed filaments are calibrated
against their \nT counterparts.

The uncertainties introduced by the filament colour propagation scheme can be of two types:
\begin{enumerate}
\item not attributing any colour to a valid filament cell, and; \label{first_point}
\item attributing the wrong colour to a filament cell. \label{second_point}
\end{enumerate}
The relative importance of uncertainties of type (i)
was assessed by recording the number of filament cells
without a colour upon application of the propagation method, as a fraction of the total number of filament
cells found by applying the \nT criteria, for different grid sizes at $z\geq3$ in the \nutco run.
These uncertainties were measured to be below the percent level at all times.

Type (ii) uncertainties are now examined.  
Since the aim of this paper is to quantify the amount of angular momentum transferred to the disk by filaments,
it is key that the trajectories of these cold flows in 
the disk's vicinity are reconstructed to a high level of precision. The filament
reconstruction stage (step $3$ of the 
colour propagation algorithm) was hence repeated over the $0.15r_{\mathrm{vir}}$ region
to ensure that the angular momentum calculations around the disk were performed
at the resolution limit (this particular choice of grid radius is justified in Section \ref{shell_section}). 
Once constructed, it is natural to wonder whether
the filament trajectories are representative of their `true' counterparts in the simulation.
A rigorous quantification of the relative discrepancies has proven elusive,  
so we provide the reader with a powerful qualitative diagnostic instead: a computer assisted visual check 
where the filament density field identified by the tracer colouring 
algorithm is compared to calculated flow lines.
Fig.~\ref{individual_filaments} shows $3$D images of the density of each filament
from the right panel of Fig.~\ref{filaments_visualization}, 
as found by implementing the colour propagation technique
on $\mathcal{G}_{0.5}$ with spacing $\Delta x\sim12\,$pc at $z\sim8$.
By zooming into the larger $0.5r_{\mathrm{vir}}$ region at the resolution limit, it is possible
to test the algorithm's ability to follow each filament onto the disk.
The viewing angle in each panel of Fig.~\ref{individual_filaments} 
has been rotated to demonstrate the nature of the individual trajectories,
and varies from filament to filament.
By comparison with the right panel of Fig.~\ref{filaments_visualization},
it is clear from Fig.~\ref{individual_filaments} that the separate filament paths are correctly identified to the point of disk contact,
and that the curvature about the disk region is particularly well-resolved,
which is most evident for the yellow filament in region $2$. 
Note that the sudden upturn of the red filament's inner trajectory in Fig.~\ref{filaments_visualization}
after mixing in the central region is also captured by the colour propagation routine. 
Thus Fig.~\ref{individual_filaments} strongly suggests that the algorithm is able to successfully construct  
the trajectories of individual filaments even though, contrary to type (i) uncertainties, it is difficult to argue that it 
does so at the percent level accuracy.

\section{Pinpointing the satellites}\label{satellite_appendix}

The details of the satellite-finding algorithm introduced in Section \ref{satellite_section}
are provided in this Appendix.

In this study, a substructure $i$ that is classed as a satellite of the host 
at a given time output $t_i$ satisfies both of the following criteria:
\begin{itemize}
\item it partially or fully infringes the host's virial sphere at $t_i$, and;
\item it has previously been a halo.
\end{itemize}
The former condition ensures that only substructures
within the host's virial region are considered,
and is hence satisfied should any region within
satellite $i$'s virial radius also be contained by the
host's virial sphere. If it is, then $i$
is flagged as a satellite candidate, otherwise it is skipped.
Assuming $i$ is a satellite candidate, 
the latter condition involves computing its merger tree, and if 
at any time output along its main branch one of its main
progenitors becomes a halo, then $i$ is flagged as a satellite.

In order to compute the main branch along $i$'s merger tree,
the progenitor that donates the most mass to $i$ is identified:
this progenitor, $j$, is called the `main progenitor',
and the link from $i$ to $j$ is referred to as the `main branch' (following \citealt{Tillson11}).
The satellite-finding algorithm imposes
the condition that in order for a subhalo to be a satellite at time $t$,
it must have been a field halo at some point earlier in its merger history. 
Note that this method allows objects to share the same main progenitor and hence 
maximizes the number of possible satellites,
resulting in the purest filamentary angular momentum signals 
once the central satellite galaxy regions have been masked according to
the prescription in Section \ref{satellite_section}.

\bsp
\label{lastpage}
\end{document}